\documentclass[11pt]{article} 
\usepackage{graphicx}%
\usepackage{multirow}%
\usepackage{amsmath,amssymb,amsfonts}%
\usepackage{amsthm}%
\usepackage{mathrsfs}%
\usepackage[title]{appendix}%
\usepackage{xcolor}%
\usepackage{textcomp}%
\usepackage{manyfoot}%
\usepackage{booktabs}%
\usepackage{algorithm}%
\usepackage{algorithmicx}%
\usepackage{authblk}

\usepackage{algpseudocode}%
\usepackage{listings}%
\usepackage{xr}

\usepackage{geometry} 
\theoremstyle{thmstyleone}%

\theoremstyle{thmstyletwo}%

\theoremstyle{thmstylethree}%

\usepackage{amssymb,amsmath,amstext,amsthm,mathtools}
\usepackage{url}
\usepackage{graphicx}
\usepackage{epstopdf}
\usepackage{color}
\usepackage{bm}
\usepackage{appendix}
\usepackage[T1]{fontenc}
\usepackage{bbold}
\usepackage{bbm}
\usepackage{latexsym}
\usepackage{float}
\usepackage{verbatim}
\usepackage{lipsum}
\usepackage{braket}
\usepackage{siunitx}
\usepackage[numbers]{natbib}
\usepackage{hyperref}
\usepackage{xcolor}

\renewcommand{\l}{\left}
\renewcommand{\r}{\right}

\makeatletter
\newcommand*{\addFileDependency}[1]{
\typeout{(#1)}
\@addtofilelist{#1}
\IfFileExists{#1}{}{\typeout{No file #1.}}
}\makeatother

\newcommand*{\myexternaldocument}[1]{%
\externaldocument{#1}%
\addFileDependency{#1.tex}%
\addFileDependency{#1.aux}%
}

\myexternaldocument{sup_arxiv}

\begin{document}

\title{Data-Driven Modeling of Dislocation Mobility from Atomistics using Physics-Informed Machine Learning}

\author[1]{Yifeng Tian}

\author[2]{Soumendu Bagchi}

\author[3]{Liam Myhill}

\author[4]{Giacomo Po}

\author[3]{Enrique Martinez}

\author[1]{Yen Ting Lin}

\author[2]{Nithin Mathew}

\author[2]{Danny Perez}

\affil[1]{Information Sciences Group, Computer, Computational and Statistical Sciences Division (CCS-3), Los Alamos National Laboratory, Los Alamos, 87545, NM, USA}

\affil[2]{Physics and Chemistry of Materials, Theoretical Division (T-1), Los Alamos National Laboratory, Los Alamos, 87545, NM, USA}

\affil[3]{Department of Materials Science and Engineering, Clemson University, Clemson, 29623, SC, USA}

\affil[4]{Department of Mechanical and Aerospace Engineering, University of Miami, Miami, 33146, FL, USA}
\date{}
\maketitle
\begin{abstract}
Dislocation mobility, which dictates the response of dislocations to an applied stress, is a fundamental property of crystalline materials that governs the evolution of plastic deformation. Traditional approaches for deriving mobility laws rely on phenomenological models of the underlying physics, whose free parameters are in turn fitted to a small number of intuition-driven atomic scale simulations under varying conditions of temperature and stress. This tedious and time-consuming approach becomes particularly cumbersome for materials with complex dependencies on stress, temperature, and local environment, such as body-centered cubic crystals (BCC) metals and alloys. In this paper, we present a novel, uncertainty quantification-driven active learning paradigm for learning dislocation mobility laws from automated high-throughput large-scale molecular dynamics simulations, using Graph Neural Networks (GNN) with a physics-informed architecture. We demonstrate that this Physics-informed Graph Neural Network (PI-GNN) framework captures the underlying physics more accurately compared to existing phenomenological mobility laws in BCC metals.
\end{abstract}




\makeatletter\@input{xy.tex}\makeatother

\section{Introduction}
Almost a century ago \cite{Orowan, Polanyi, doi:10.1098/rspa.1934.0106}, dislocations were identified as carriers of plastic deformation in crystalline materials. A key construct of the theory is the response of dislocations to an applied stress state, the \emph{dislocation mobility law}, which describes the relationship between the force experienced by the dislocation ($\mathbf{f}$) and the resultant velocity ($\mathbf{v}$). In crystals with low lattice resistance and planar core structure, e.g., Face-Centered Cubic (FCC) metals, the mobility law is usually simplified to a viscous damping relationship $\mathbf{v}=\mathbf{M}(T)\cdot\mathbf{f}$, where $\mathbf{M}$ is the matrix of mobility coefficients and $T$ is the temperature \cite{HULL201143}, under the assumption of over-damped dynamics. In metals, $\mathbf{M}$ encodes dissipation mechanisms due to both electrons and phonons, with dominant phonon contribution except at very low temperatures. Theoretical calculations of phonon scattering predict a linear dependence of the mobility coefficients on temperature \cite{Brailsford19721380} with a temperature-independent leading term proposed recently \cite{Swinburne2014, Swinburne2015}. While this picture holds for edge dislocations in Body-Centered Cubic (BCC) metals, the complicated core structure of the BCC screw dislocations results in additional complexity. Screw dislocations in BCC are known to have high lattice resistance. This in turn results in a strong temperature dependence of the flow stress as the mobility of screw dislocations in BCC are controlled by kink-pair nucleation and migration enthalpy. In addition, screw dislocation motion is strongly influenced by multiple stress components, in addition to the resolved shear stress in the direction of the Burgers vector, resulting in non-Schmid effects \cite{Christian,GROGER20085401,hale2015nonschmid}. These complexities are typically accounted for in the mobility law by introducing functional dependencies on the local stress tensor $\boldsymbol{\sigma}$, line orientation $\bm\xi$, temperature and possibly other local internal variables, that is \cite{HirthLothe}
\begin{align}
\bm v =\bm g_\alpha(\bm \sigma,\bm\xi, T,\ldots)\, .
\label{moblilityLocal}
\end{align}
The function $\bm g_\alpha$ is known as the (glide) \emph{mobility law} for slip system $\alpha$ \cite{10.1063/1.1728907, Christian1970307, nabarro1967theory} The functional dependencies of the mobility law are in turn informed by phenomenological approaches based on kink-pair nucleation and kink diffusion. A review of the phenomenological approach for deriving dislocation mobility laws can be found in \cite{PO2016123}. State-of-the art methods to develop such models for dislocation mobility (e.g., in BCC materials) \cite{gilbert2011stress,queyreau2011edge,maresca2018screw} have generally focused on screw and edge type dislocations and smooth interpolation between these for other line orientations. However, this approach presents multiple pitfalls. For example, it has been shown previously that ``singular'' line orientations exist in BCC metals \cite{doi:10.1073/pnas.1206079109, Yamaguchi_1973} which result in cusps in the mobility. Dislocations in these singular orientations, and those vicinal to them, move by motion and propagation of kink pairs. Therefore, their mobility cannot be accurately represented using a smooth interpolation between screw and edge dislocations.  Moreover, the mobility becomes a complex function of local chemical environments in the case of conventional and high entropy alloys (HEAs). The atomically-rough energy landscape for dislocation motion, which is believed to be an important characteristic of HEAs \cite{varvenne2016theory, rida2022influence}, results in  Arrhenius-type mobility laws which are highly non-linear in stress and temperature \cite{PO2016123}. It has been demonstrated in BCC-based HEAs that the mobility of mixed dislocations \cite{chen2020unusual} and multiplicity of glide planes \cite{wang2020multiplicity} can drive the overall plastic behavior, in contrast to BCC metals. The traditional phenomenological approach informed by a handful of intuition-driven data-points to derive the mobility law can quickly become error-prone and cumbersome in such scenarios. 

These challenges can in principle be addressed by using data-driven approaches, whereby active learning techniques are used to efficiently explore the parameter space for learning the dislocation mobility law with minimal human intervention. Such approaches are especially relevant with the advent of exascale computing platforms where hierarchical multi-scale methods, deployed at scale with automated model refinement, can potentially be used to drive materials design for targeted functionality. Moreover, the improved accuracy of Machine-Learning (ML) interatomic potentials \cite{montes2022training} promises affordable near-quantum accurate predictions, which could enable upscaling of atomic-level defect dynamics to predictive mesoscale models such as Discrete Dislocation Dynamics (DDD) \cite{LeSar2020, GiacomoReview, ZHOU20101565, Zbib, doi:10.1080/14786430801992850, CAI2006561}. To date, the exploration of these ideas is in a relatively nascent stage. Previous studies have explored the possibility of using graph-theoretical techniques for learning specific aspects of dislocation glide. Moraes et al.~\cite{BARROSDEMORAES2021110569} developed a graph-based surrogate model for glide of edge dislocations in BCC Fe, trained to molecular dynamics (MD) data. In this representation, sub-domains within the MD model are represented using nodes on a ring graph. The dislocations are then represented as random walkers which jump between neighboring nodes following a Poisson process. More recently, Bertin and Zhou \cite{BERTIN2023112180} developed a Graph Neural Network (GNN) approach to model DDD, with the aim of replacing its time-integration procedure. They demonstrated that this DDD-GNN approach can learn ground-truth (GT) DDD data of a dislocation gliding in a field of obstacles. While these studies show the potential for accelerating materials modeling through machine learning, bridging mesoscale dynamics (i.e. DDD) with information from higher fidelity (i.e. atomistic) simulations involving more general defect characteristics (i.e. mixed orientations of dislocations) remains to be explored.

\begin{figure}[htb] 
    \centering
    \includegraphics[trim=10 20 0 0,width=0.9\textwidth]{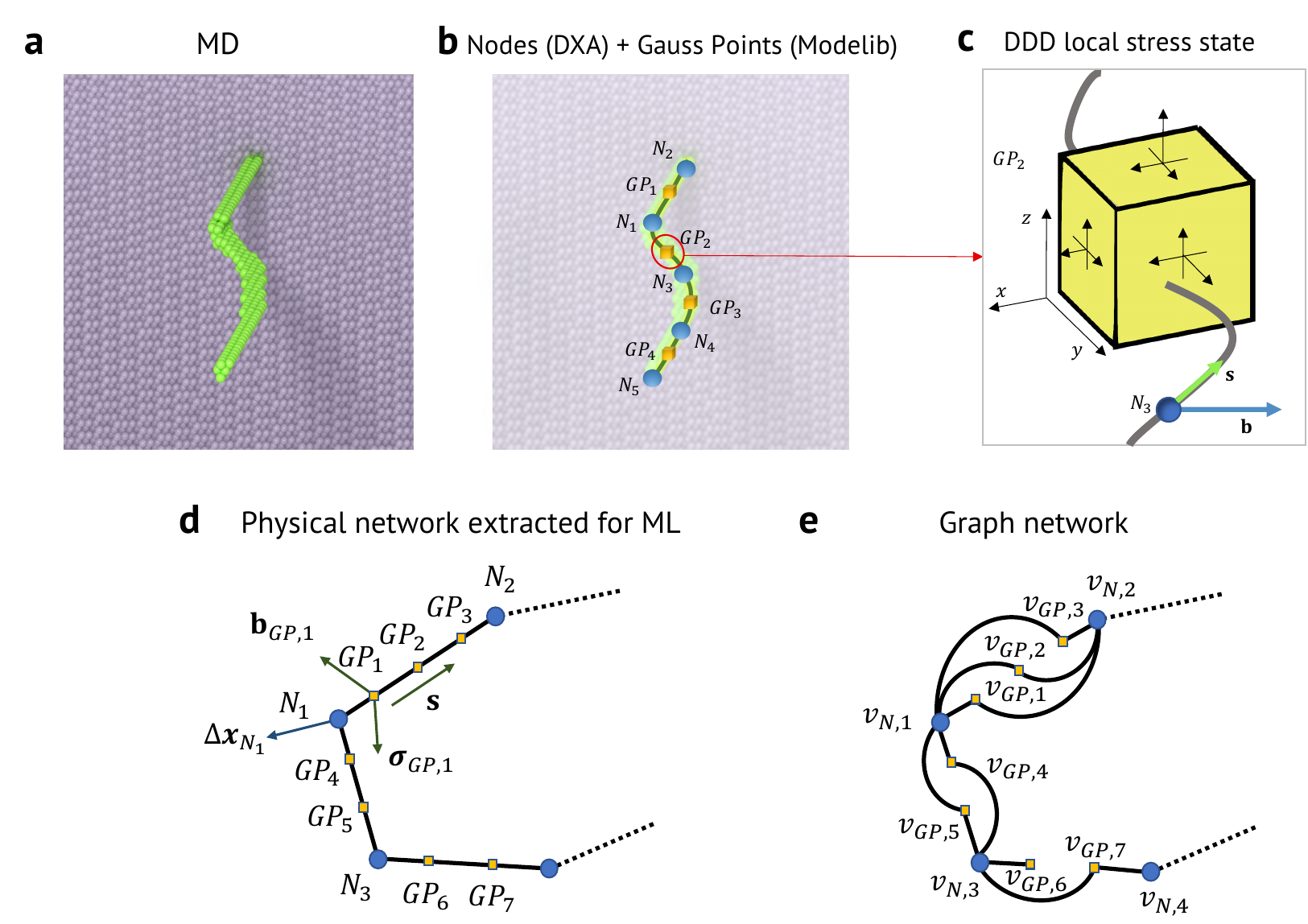}
    \caption{\textbf{Schematic of ground truth data generation:} \textbf{a} Atomistic representation of a dislocation segment (green atoms represent the defect core while the purple denotes pristine atoms). \textbf{b} Idealised nodal representation of the dislocation segment obtained from Dislocation Extraction Algorithm (DXA). Blue spheres represent dislocation nodes (denoted as $N_1, N_2, ...$) and yellow cubes represent the Cauchy stress tensor, one of the features, defined at Gauss Points (denoted as $GP_1, GP_2, ...$) using the linear elasticity kernel in MoDELib Discrete Dislocation Dynamics (DDD) code. \textbf{c} Zoom-in of the local stress state ($\bf \sigma$) at a given Gauss point, with the definition of segment tangent vector (${\bf s}$) and the Burger's vector ($\textbf{b}$). \textbf{d} General representation of the dislocation configuration with multiple Gauss points between nodes. \textbf{e} Graph representation of the dislocation configuration.}
    \label{fig:Schematic}
\end{figure}

Our objective in this study is to develop a generalizable framework for data-driven modeling of dislocation dynamics mobility law. Our proposed framework integrates high-throughput MD simulation datasets, physics-inspired machine learning model structures, and uncertainty quantification, accelerating the development of mesoscale models of dislocation mobility. A schematic diagram with different components of our proposed modeling workflow is shown in Fig.~\ref{fig:Schematic}. To demonstrate our approach, GT data was generated from classical MD simulations of dislocation motion using an Embedded Atom Method (EAM) type interatomic potential (Fig.~\ref{fig:Schematic}a); the workflow would remain formally identical, albeit more expensive, if ML interatomic potentials were used instead. The MD simulation data consists of dislocation dipoles and shear loops moving under different values of resolved shear stress (RSS) at various temperatures. Note that, while Non-Schmid effects are important in BCC materials, our aim here is to demonstrate the ML workflow supported by high-throughput large-scale atomistic trajectory generation protocols for a relatively simple data set with only RSS. Coarse-grained representations of dislocations were obtained from this atomistic representation using the Dislocation Extraction Algorithm (DXA) as implemented in the Open Visualisation Tool (OVITO)\cite{ovito} software. The DXA representation, which consists of nodes with line segments between them, was converted into a numerical representation with nodes, quadrature points or Gauss points (GPs), and line segments as shown in Fig.~\ref{fig:Schematic}b. Relevant input features of these dislocation configurations, which will serve as input for an ML procedure, such as the Cauchy stress tensor ($\bf \sigma$, shown schematically in Fig.~\ref{fig:Schematic}c), segment tangent vector (${\bf s}$), and the Burger's vector ($\textbf{b}$) were computed using a recent version of the open-source Mechanics of Defects Evolution Library (MoDELib) supporting periodic boundary conditions \cite{pachaury2022discrete}. We remark that the computation of stress tensor by MoDELib is based on linear elasticity theory and the resulting stress can deviate from the true stress experienced by the dislocation. Nevertheless, we hypothesize that the stress tensor computed by MoDELib contains sufficient information for us to make accurate predictions, and the error due to the linear assumption, if any, will be corrected by our proposed GNN-based ML procedure below. We also note that the physical features of the dislocation lines, such as the Burger's vector, segment tangent vector, and local stress, are defined on both the GPs and nodes, while the displacements ($\Delta$${\bf x}$) are defined on the nodes alone, with several GPs being present between any given pair of nodes in general. Dislocation network, thus represented as a set of nodes and Gauss points (GPs) in DDD can be naturally encoded as a graph. In Figures \ref{fig:Schematic}d and \ref{fig:Schematic}e, we present the visualization of the dislocation network, showcasing both its physical representation extracted from DDD for ML and its graph representation. The physical network serves as an abstract depiction of the dislocation network, which consists of nodes ($N_i$) and GPs ($GP_i$), connected by the dislocation segments. Figure \ref{fig:Schematic}e illustrates the corresponding heterogeneous graph, a type of graph containing vertices with different features, for learning purposes. Within this heterogeneous graph, we introduce two types of vertices, $v_N$ representing the nodes and $v_\text{GP}$ representing the GPs in the physical network. In the graph representation, instead of connecting $v_\text{GP}$'s sequentially into a segment between nodes $v_N$, we directly link GPs to their nearest nodes. This heterogeneous graph enables the direct exchange of information between GPs and their neighboring nodes, facilitating efficient information transfer. 

To achieve the goal of the proposed work, we present a general physics-informed data-driven approach for modeling dislocation mobility law. Our approach incorporates a physics-inspired mathematical structure based on the theoretical descriptions of dislocation mobility into the proposed GNN model. Features are designed to respect physical constraints, e.g.~rotational and translational invariance. The proposed physics-informed GNN mobility law (PI-GNN) is trained using a coarse-grained MD dataset, with a loss function defined in Sec.~\ref{sec:PIGNN}. To quantify the predictive uncertainty and facilitate accelerated learning, we employ an uncertainty quantification-driven active learning framework \cite{kroghNeuralNetworkEnsembles} that is capable of providing on-the-fly uncertainty measures and automatically query new MD simulations to improve the model's predictability and uncertainty. By studying the mobility of $\{110\}$ dislocations in BCC Fe, we demonstrate the effectiveness of this PI-GNN framework to accurately capture the relevant physics of dislocations systematically.

\section{Results}
\label{sec:results}
\textit{Learning the mobility function using Graph Neural Networks \cite{bronstein2017geometric} in the Physics-informed Machine Learning (PIML) framework.} A GNN is a type of neural network designed to act on graph-structured data, where nodes represent entities (e.g., the nodes and GPs on the dislocation) and edges represent relationships (e.g., the spatial proximity of the nodes) between them. GNNs learn powerful representations of graph data by propagating information between neighboring nodes. Unlike convolutional neural networks on grids, GNNs respect rotational and translational invariance, making them well-suited for learning on irregular graph domains. As such, GNNs serve as an ideal neural architecture for the coarse-grained representation of dislocation structures generated by DXA. Furthermore, the design of the PI-GNN architecture was informed by established physics in linear elasticity theory pertaining to dislocation dynamics in continuum media \cite{PO2016123}, i.e., 
\begin{eqnarray}
\label{eqn:DDDmodel}
\frac{{\rm d} {\bf x}}{{\rm d} t} = \frac{{\bf f}( {\bf \sigma}, {\bf b}, {\bf s})}{B(T, {\bf b}, {\bf s})} \exp \l (-\frac{\epsilon({\bf \sigma}, {\bf b}, {\bf s},T)}{k_B T}\r),
\end{eqnarray}
where ${\bf f}$, $B$, and $\epsilon$ are separate networks modeling the contribution of the Peach-Koehler force, drag, and thermally activated processes in dislocation dynamics. These networks, referred to as F-net, B-net, and E-net, are jointly trained on time-series data of the dislocation dynamics, generated by MD and processed through the DXA-MoDELib pipeline. Through message passing \cite{bronstein2021geometric}, the PI-GNN not only uses the local information but also information on nearby nodes to learn higher-order corrections to linear theories. 
More details of the PI-GNN model we adopted can be found in Sec.~\ref{sec:Methods} and in the Supplementary Sec.~\ref{sisec:PI-GNN}.\\

\begin{figure}[htb]
    \centering
    \includegraphics[width=0.75\textwidth]{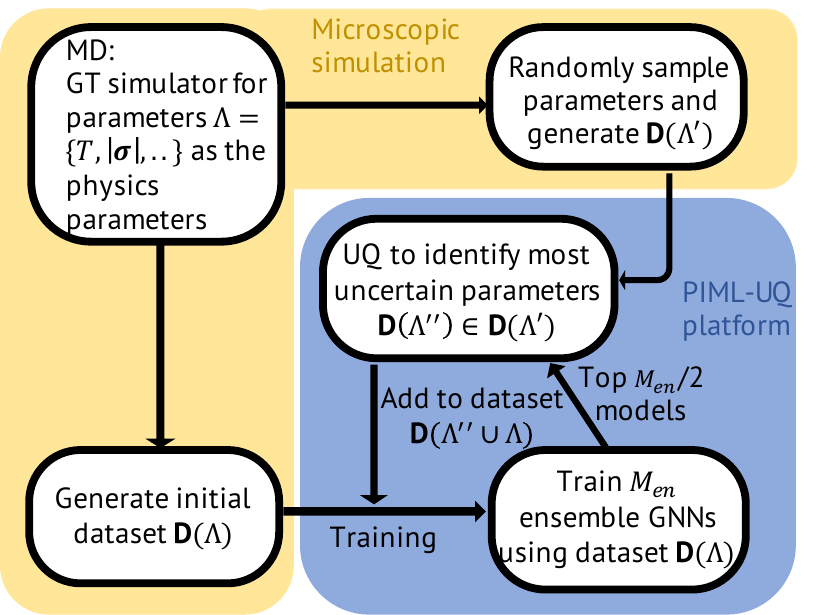}
    \caption{\textbf{Flow chart of UQ-driven active learning framework.} The UQ-AL framework is composed of two computational platforms, the high-throughput microscopic MD simulations, and PIML-UQ platform. The UQ-AL workflow begins by generating and learning from an initial dataset, and then progressively queries new datasets where the trained models exhibit the most  uncertainty. In each iteration, we independently train an ensemble of identical GNN models with randomized initialization of weights and biases. These models are then evaluated on newly generated datasets with randomly sampled simulation parameters, and the uncertainties of the models’ single-step predictions are quantified. We integrate the top 50\% of new configurations that induce the highest predictive uncertainty into the existing training dataset. The PI-GNN ensemble is retrained on this augmented dataset, and the process repeats.}
    \label{fig:MDALLoopsummary}
\end{figure}

\noindent \textit{Uncertainty quantification-driven active learning (UQ-AL) framework to facilitate efficient learning.} We propose a workflow, outlined in Fig.~\ref{fig:MDALLoopsummary}, for training the PI-GNN while aiming to minimize its predictive uncertainty. Rather than generating one large dataset all at once, the workflow begins by learning from an initial dataset, and then progressively queries new datasets where the trained models exhibit the most uncertainty. In each iteration, we independently train $M_{en}$ identical GNN models with randomized initialization of weights and biases. These models are then evaluated along the trajectories of a set (8) of newly generated datasets with randomly sampled simulation parameters $\Lambda=\{T, \vert \bm \sigma\vert,...\}$, and the uncertainties of the models' single-step predictions are quantified. This can be considered as an ensemble learning \cite{kroghNeuralNetworkEnsembles} method. We integrate the top 50\% of new configurations that induce the highest predictive uncertainty into the existing training dataset. The PI-GNN ensemble is retrained on this augmented dataset, and the process repeats. We note that this method is an approximation of more rigorous but computationally expensive UQ-AL frameworks (e.g., Bayesian Optimization \cite{garnettBayesianOptimization2023}) for two reasons. First, the variability of the $M_{en}$ model predictions is only a lower bound to the true uncertainty \cite{kroghNeuralNetworkEnsembles}. Second, we select the most uncertain candidates from randomly sampled simulation parameters, rather than directly querying the single most uncertain point across the entire simulation parameter space. More details of the proposed UQ-AL framework can be found in Sec.~\ref{sec:Methods}.

We illustrate the advantages of the proposed UQ-AL framework in learning the mobility law using the dipole dataset generated from MD simulations as described in Sec. \ref{sec:MD}. In this numerical experiment, the PI-GNN models are continuously retrained as more data are acquired and integrated into the training set, and we compare the performance of the model when the new datasets are either actively selected by the UQ-AL or passively queried by randomization.
We fix the total number of ensemble models $M_{en}$ at 20. The initial dataset ${\bf D}(\Lambda)$ are generated on the set of simulation parameters: temperature $T \in \{300\text{ K}, 500\text{ K}\}$, RSS $\vert {\bf \sigma} \vert \in \{ 0.25 \text{ GPa}, 0.5 \text{ GPa} \}$, and the line orientation (i.e., the angle between the tangent vector ${\bf s}$ and the Burger's vector ${\bf b}$) $\text{angle}({\bf s}, {\bf b}) \in \{ 45^{\circ}, 135^{\circ} \}$. In each iteration, we maintain a candidate pool of 8 new trajectories by sampling the simulation parameters, $T\sim \text{Unif}(100\text{ K}, 1000\text{ K})$, $\vert {\bf \sigma} \vert \sim \text{Unif}(0.1 \text{ GPa}, 1.0 \text{ GPa})$, and $\text{angle}({\bf s}, {\bf b})\in (0^\circ, 180^\circ) $, to create new trajectories of dislocation dynamics. With UQ-AL, the trained PI-GNN ensemble is evaluated along the trajectories to quantify their averaged single-step predictive uncertainty; we select and integrate the top 50\% (4) trajectories with the highest predictive uncertainties (i.e, the PI-GNN ensemble models disagree with the most) into the existing training dataset. 
With passive learning, we randomly select 4 new trajectories from the pool and integrate them into the training dataset.
For both methods, the selected trajectories are removed from the candidate pool, which is then replenished by 4 newly sampled parameter settings to generate the trajectories for the next iteration. 

\begin{figure}[htb]
    \centering
    \includegraphics[width=0.9\textwidth]{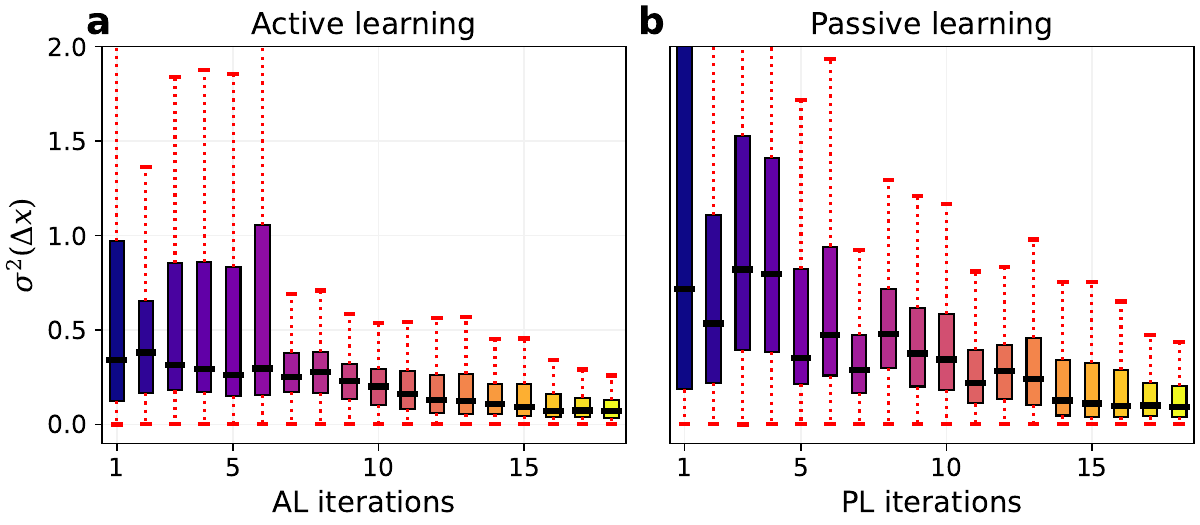}
    \caption{\textbf{Comparison of the proposed UQ-AL framework with a passive learning method}. The uncertainty measure of predictions for nodal displacement as a function of the \textbf{a} active (AL) and \textbf{b} passive learning (PL) iterations.}
    \label{fig:ALPL}
\end{figure}

Figure \ref{fig:ALPL} illustrates how the model performance improves as more data are acquired into the training set, for both active and passive selection of the datasets. Evidently, with the proposed UQ-AL framework, the prediction uncertainty is consistently reduced through the iterations. In comparison, the predictive uncertainty with passive selection is higher and fluctuates more between iterations, even though it eventually decreases in the large-data limit. Suppose we set the uncertainty threshold to be $25\%$ of the mean nodal displacement, $\sigma(\Delta {\bf x}) = 25\% \mu(\Delta {\bf x}) $, the active and passive acquisition of the new training dataset requires 8 and 15 iterations, or equivalently 36 and 64 queries (initial 8 fixed, then 8 randomly sampled ones to populate the candidate pool in the first iteration, and 4 randomly sampled ones in each of the subsequent iterations) of new trajectories, respectively. As such, UQ-AL saves more than $40 \%$ of computational resources for generating training dataset in this experiment. Because the proposed UQ-AL can efficiently query computationally expensive high-fidelity datasets to learn mesoscale models and thus accelerate the development of data-driven mesosocpic models for DDD,  we deploy this framework in the following analysis as part of the data acquisition pipeline. \\

\begin{figure}[htb]
    \centering
    \includegraphics[trim=175 150 140 150, width=0.9\textwidth]{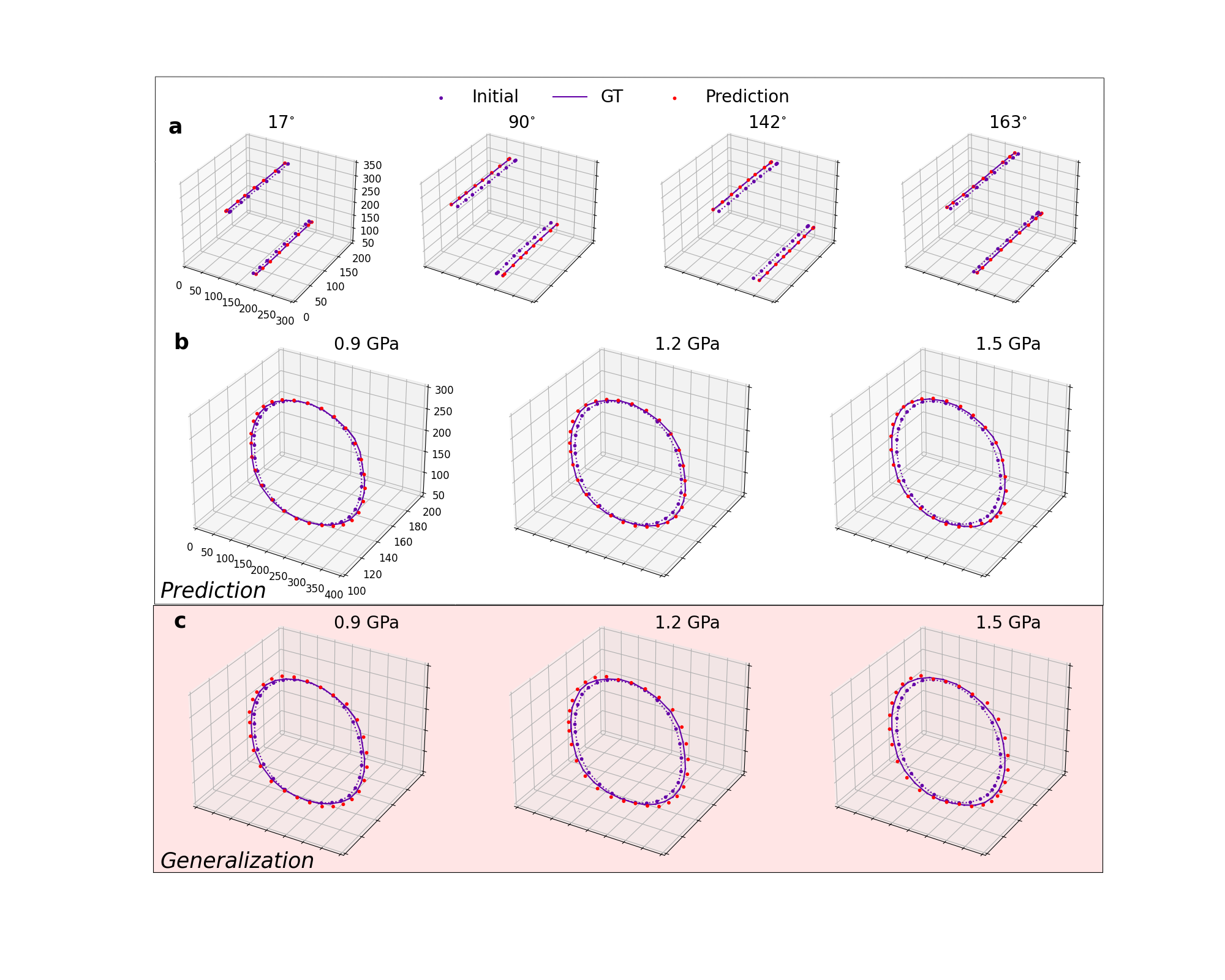}
    \caption{\textbf{The predicted single-step evolution for various dislocation configurations and learning method.} \textbf{a} PI-GNN mobility law is trained on coarse-grained dipole dislocation dataset with different RSS values, temperatures, and angles between Burger's vector and dipole tangent. The predictions are also made on dipole configurations with different orientations at 500K and 0.9GPa. \textbf{b} PI-GNN mobility law trained on dislocation loop expansion data and prediction on dislocation loops at 500K and 0.9, 1.2, and 1.5GPa. \textbf{c} Testing the generalizability of the PI-GNN mobility law by training on dipole dislocation data and predicting on loop expansion.}
    \label{fig:Evolution}
\end{figure}

\noindent{\textit{Predictability and generalizability of the PI-GNN mobility law.}} In the following analysis, the proposed UQ-AL is used to train the PI-GNN model until the predicted uncertainty is reduced to 25\% of the mean nodal displacement. The 50\% top-performing models in the ensemble are selected for the following validation analysis, in which we compare the single-step ($\Delta t \approx 1ps$) prediction of the trained models to a validation dataset. See Sec. \ref{sec:Methods} for details on the train/test/validation dataset. We designed three different experiments: (a) the PI-GNNs ensemble is trained on a dataset comprised of nominally straight dislocation dipoles and compared against the dipole validation dataset, (b) the PI-GNNs ensemble is trained on a dataset comprised of expanding shear loops and tested on a loop validation dataset, and (c) the PI-GNNs ensemble is trained on the dipole dataset and tested on the loop dataset. Figure \ref{fig:Evolution} shows selected examples of the single-step prediction; more results can be found in Supplementary Sec.~\ref{sisec:pignnresults}

We observe that the PI-GNN is capable of learning an effective mobility law for various dislocation dipole configurations. Experiment (a) (Fig.~\ref{fig:Evolution}a) shows that the PI-GNN model trained on various conditions of temperature, RSS, and line orientation of a dipole configuration, can learn an effective mobility law to predict the evolution of the dislocation with an out-of-sample dipole configuration. In contrast to experiment (a), where we explicitly sampled the line orientation of the dislocation for training, the PI-GNN learns the dependence of the line orientation directly from the angular distribution on the evolving loops in experiment (b), and accurately predicts the evolution of out-of-sample loop configurations. Experiment (c) shows that the PI-GNN is able to learn the mobility law from the dipole dataset and generalize to the out-of-distribution prediction task for loop configurations. Such generalizability indicates that the PI-GNN is learning the physics of the dislocation dynamics, instead of memorizing the training data.

\begin{figure}[htb]
    \centering
    \includegraphics[width=0.96\textwidth]{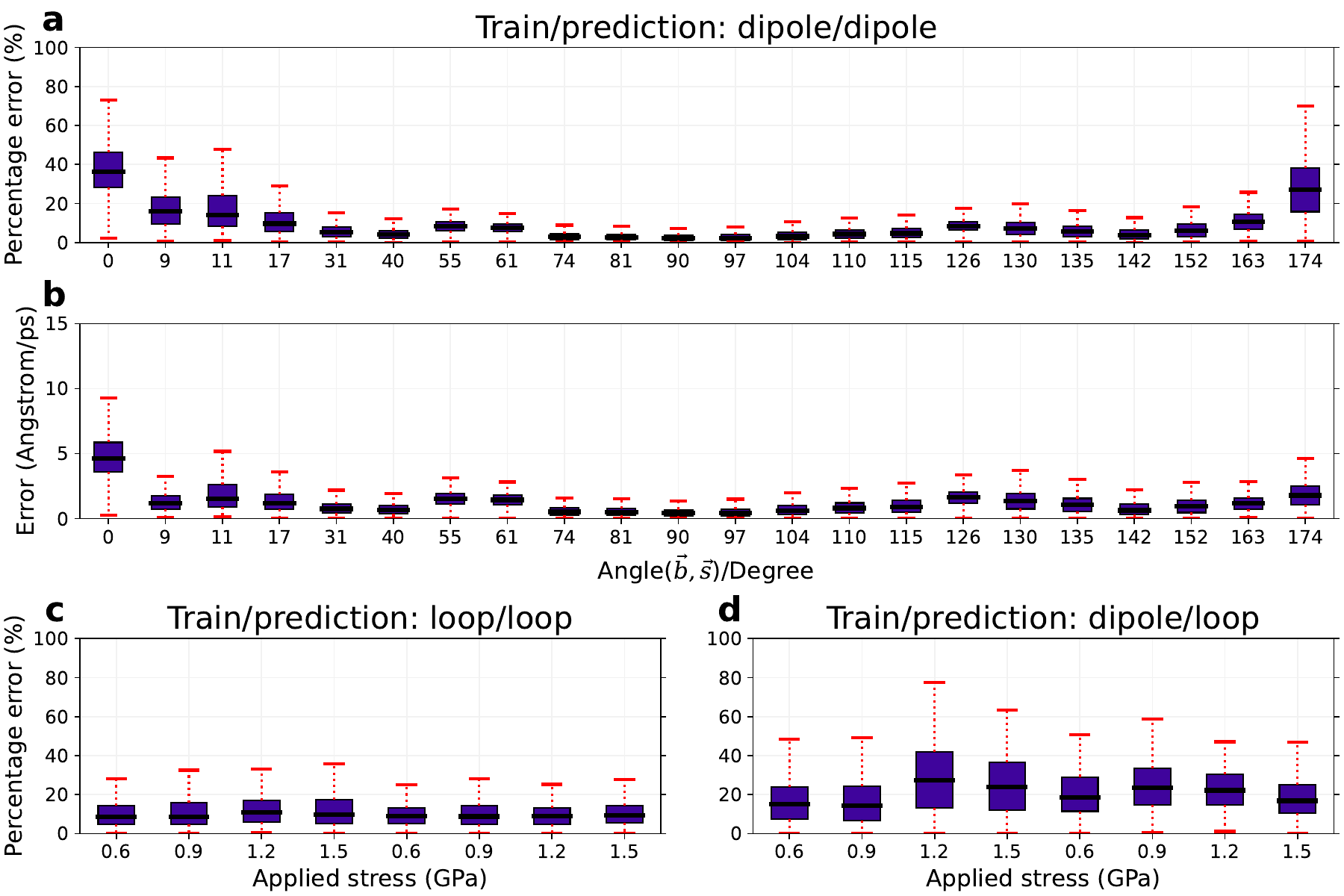}
    \caption{\textbf{The predicted single-step error for various dislocation configurations and learning method}. PI-GNN mobility law is trained on coarse-grained dipole dislocation datasets with different applied stress levels, temperatures, and angles between Burger's vector and dipole tangent. The predictions are also made on dipole configurations with different orientations at 500K and 0.9GPa. Both the \textbf{a} percentage error and \textbf{b} absolute unnormalized error are shown. \textbf{c} PI-GNN mobility law trained on dislocation loop expansion data and prediction on dislocation loops at 500K and 700K and various applied stress levels.  \textbf{d} Testing the generalizability of the PI-GNN mobility law by training on dipole dislocation data and predicting on loop expansion.}
    \label{fig:l2Evolution}
\end{figure}

Quantitative error analyses of the learned PI-GNN mobility law are reported in Fig.~\ref{fig:l2Evolution}, where the panels' layout is analogous to Fig.~\ref{fig:Evolution}. The prediction error is quantified by both the percentage and absolute errors, the computation of which is based on the proposed loss function (See Sec.~\ref{sec:PIGNN} for more information). For the dislocation dipole, the median prediction error is in general very small for different orientations, temperatures and stress levels (500K and 0.9GPa is shown in the figure). For screw-like dislocation where the angle(${\bf b}, {\bf s}$) is close to $0^\circ$ and $180^\circ$, the percentage error is comparatively larger. {This is mainly caused by an imbalanced training dataset, due to the timescale separation between the dynamics of screw dislocations and mixed/edge dislocations. At the timescales of the dataset (several picoseconds) used in this study, the screw dislocation dynamics are dominated by thermal fluctuations. Therefore, the signal-to-noise ratio of the nodal displacement, which is the prediction goal of the PI-GNN mobility law, is much smaller than that of the edge. The small signal-to-noise ratio can only be seen for configurations near the pure screw configuration (i.e., angle$\left(\mathbf{b}, \mathbf{s}\right)=0^\circ$ and $180^\circ$), while the majority of the data points do not have this signature. Consequently, the trained PI-GNN model predicts the edge and mixed configurations better than the pure screw configuration. Future work to rectify the training data imbalance will be discussed in Sec.~\ref{sec:discussion}.}
When we compare the absolute error in Fig.~\ref{fig:l2Evolution}b, the absolute errors for different angles are at approximately the same range. For dislocation loops, the single-step prediction percentage error is also very small. The errors in generalization are plotted in Fig.~\ref{fig:l2Evolution}d. As explained above, we train the model in this case using dislocation dipoles and make predictions on shear loops. Compared to Fig.~$\ref{fig:l2Evolution}$c, even though the percentage error is larger, it still falls in an acceptable range, with the median at approximately $20\%$. This shows that with the PIML structure and features, the PI-GNN model can still capture roughly $80\%$ of the total motion even without any additional training on the complex dataset. These numerical experiments demonstrate the potential generalizability of the proposed UQ-AL PI-GNN framework for modeling complex dislocation networks using simple datasets.\\

\begin{figure}[htb]
    \centering\includegraphics[trim=150 40 10 0, width=0.9\textwidth]{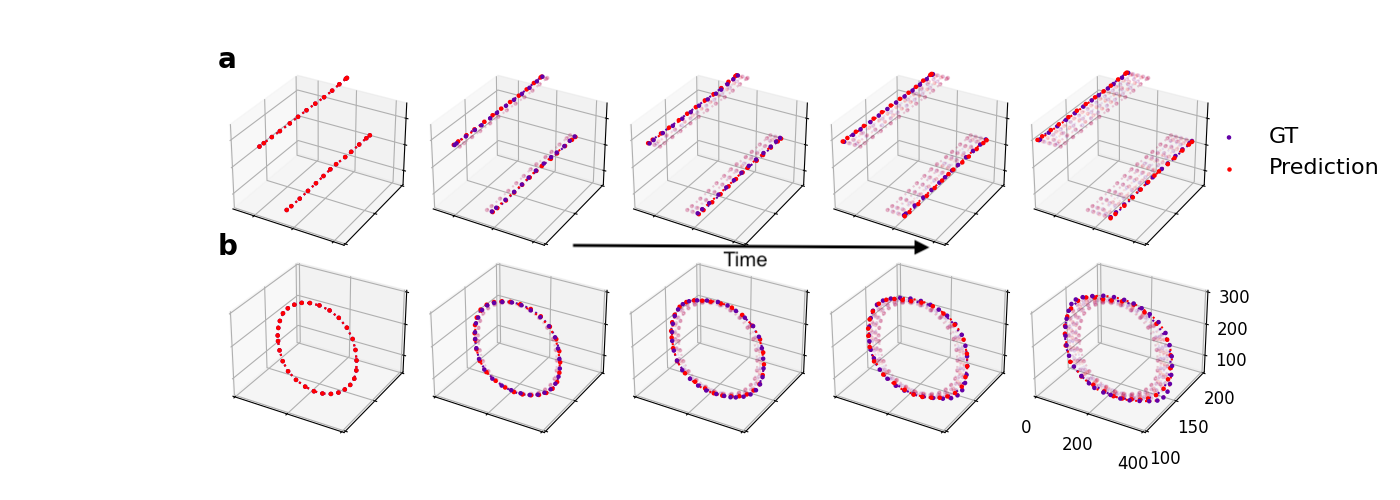}
    \caption{\textbf{The predicted multi-step evolution for various dislocation configurations and learning method}. For a given snapshot, the GT is represented using blue-filled circles, the PI-GNN prediction using red-filled circles with translucency applied to configurations from all preceding time steps. \textbf{a} A dislocation dipole(angle$({\bf b}, {\bf s}) \approx 80^{\circ}$) at 0.9GPa and 500K predicted using PI-GNN mobility law trained on dislocation dipole data. \textbf{b} An expanding dislocation loop at 500K and 1.2GPa, predicted using PI-GNN mobility law trained on dislocation loop expansion data.}
    \label{fig:EvolutionMulti}
\end{figure}

\noindent \textit{Integrating PI-GNN mobility law with DDD for iterative multi-step predictions.} In Fig.~\ref{fig:EvolutionMulti}, the predicted iterative multi-step evolution of both dipole and loop configurations are visualized at different consecutive time steps. We  show results at 0.9GPa and 500K, and angle$({\bf b}, {\bf s}) \approx 80^{\circ}$ for the dislocation dipole. For each row in Fig.~\ref{fig:EvolutionMulti}, the panels from left to right show the forward evolution in time, where configurations from preceding time steps are also visualized as translucent markers. As observed in Fig.~\ref{fig:EvolutionMulti}, the PI-GNN mobility law enables us to achieve qualitatively accurate long-time predictions. The coarse-grained model prediction will eventually deviate from the GT, but within the visualized prediction horizon, the relative error remains less than 10\%. Quantitative analysis of the errors in multi-step prediction is provided in Fig.~\ref{fig:MultistepError}. \\

\noindent \textit{Interpretability of the PI-GNN mobility law.} The phenomenological model, represented by Eq.~\ref{eqn:DDDmodel}, involves three distinct physical aspects of dislocation dynamics, namely the force experienced by the dislocation given a stress state, dissipative mechanisms resulting in drag, and thermal activation, each of these making a unique contribution. We model each aspect using a separate GNN as shown in Eq.~\ref{eqn:DDDmodel} and train them altogether with a single loss function. Even though quantitatively isolating each process from the trained PI-GNN mobility law is challenging, the proposed framework still offers qualitative insights into their sensitivity and dependence on the input features. For instance, scaling both F-net and B-net by a constant factor does not alter the output. To resolve the identifiability issue arising from the linear constant factor, we compute the ratio of F-net over B-net, as well as other components of the PI-GNN mobility law, against various input features.

\begin{figure}[htb]
    \centering
    \includegraphics[width=\textwidth]{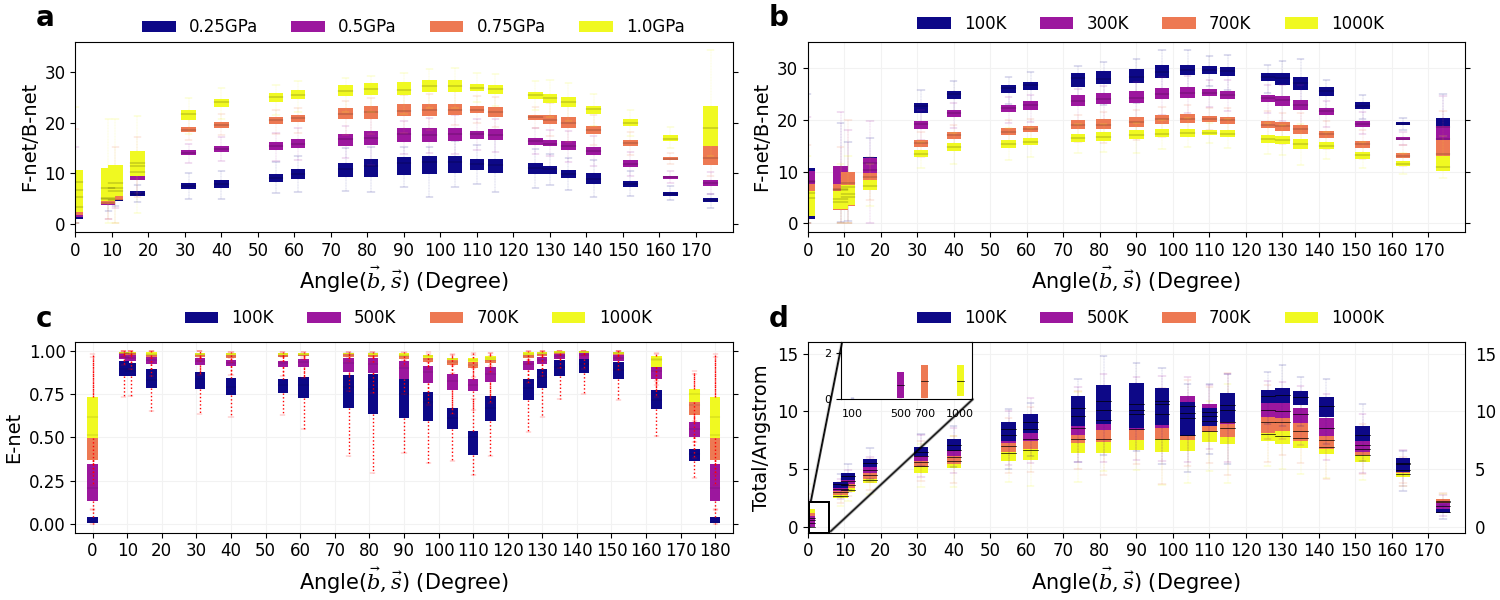}
    \caption{\textbf{Physics-informed interpretation of the three components of the PI-GNN mobility law}. \textbf{a}  Box plots of the $\frac{\mathbf{F}}{\mathbf{B}}$ for different orientations, applied stress levels at 500K. \textbf{b} Box plots of the $\frac{\mathbf{F}}{\mathbf{B}}$ for different orientations, and different temperatures at 0.5GPa.\textbf{c}  Box plots of the $\mathbf{E}$ for different orientations, and different temperatures at 0.5GPa. \textbf{d}  Box plots of the total predicted displacement for different orientations, and different temperatures at 0.25GPa.} 
    \label{fig:PIML}
\end{figure}

In Fig.~\ref{fig:PIML}a, we plot the output of F-net/B-net at different angles and applied stress levels. Despite the identifiability challenge, combining the outputs of both networks allows us to study their collective contributions to the prediction. With the increase of applied stress level, the predicted contributions to nodal displacement become larger as expected. For different angles representing edge, screw or mixed dislocations, the magnitude of the output is larger around $90^{\circ}$ (i.e., edge dislocation), and small at $0^{\circ}$ and $180^{\circ}$ ( i.e., screw dislocation), accurately capturing the well-known disparities in the mobility of screw and edge dislocations in BCC metals. An interesting observation is that the output is not symmetric with respect to $90^{\circ}$ and is skewed toward larger angles. In Fig.~\ref{fig:PIML}b, the dependence of the combined F-net/B-net on temperatures is plotted. We observe that the increase in phonon drag with increasing temperature is also captured by the GNNs. It is important to note that the temperature-dependence of phonon drag is known to be stronger for edge-like dislocations \cite{queyreau2011edge} and the GNN learns this behavior; as shown in Fig.~\ref{fig:PIML}b that the temperature-dependence is stronger close to $90^{\circ}$ or edge dislocation as it is apparent from the GT MD data (see Fig.~\ref{fig:Schew_EdgeM111Mobility} in Supplementary Information for dislocation velocities as a function of stress and temperature for screw, edge, and mixed configurations).

Figure \ref{fig:PIML}c illustrates the dependence of the contribution from E-net output on the dislocation orientations. We note that for screw dislocations, with angles$({\bf b}, {\bf s}) \approx 0^{\circ}, 180^{\circ}$, the PI-GNN mobility law qualitatively captures the comparatively higher activation energy, leading to lower output from the E-net. Conversely, at the mixed configurations, the E-net output becomes larger. In addition, we observe a cusp at $110^{\circ}$, predicting a reduced contribution from the E-net for that singular orientation. The combined output from all three networks in Fig.~\ref{fig:PIML}d shows that the PI-GNN mobility law is capable of predicting expected dependencies on temperatures and angles. For instance, the increased phonon drag from high temperatures causes the dislocation lines to move slower at some orientations. In the case of screw dislocations with a low applied stress level, the nodal displacement is dominated by thermal activation, which has a reversed order compared to other angles (see the inset in Fig.~\ref{fig:PIML}d). This is also apparent for the singular orientation at to $110^{\circ}$, where thermal activation dominates at low temperatures. Such cusps in dislocation mobility is known to be present in BCC metals \cite{doi:10.1073/pnas.1206079109, Yamaguchi_1973} due to singular dislocation orientations which move by kink nucleation and propagation. The predicted trends are aligned very well with what is observed in MD simulations (see Fig.~\ref{fig:rawMDcomMobility0.25GPa}), demonstrating the accuracy of the PI-GNN mobility law and the ability of GNNs to learn complex dependence of dislocation mobility on different parameters.

\section{Discussion}
\label{sec:discussion}
The central objective of this work is to provide a general, physics-informed data-driven framework for modeling the mobility law in discrete dislocation dynamics simulations autonomously from higher fidelity molecular dynamics data. When constructing the modeling framework, we focused on the following aspects: generalizability, interpretability, and uncertainty quantification. To achieve generalizability for the model, we adopt the flexible graph-based data structure to represent the dislocation network, which can be easily extended to dislocation configurations beyond those studied in this paper. We demonstrate that this framework is capable of learning many complex aspects of dislocation motion in BCC metals, such as highly anisotropic dislocation mobility between screw and edge configurations, cusps in dislocation mobility at singular line orientations, and temperature dependence of phonon drag. The interpretability of the proposed framework results from the physics-inspired structure of the PI-GNN mobility law and physics-informed feature engineering. We illustrate the interpretability of the proposed framework by analyzing learned components of the PI-GNN, which showed qualitative agreement with various physical aspects of dislocation mobility. Uncertainty quantification is built into the proposed framework to inform data selection from MD simulations for active learning, which is shown to significantly accelerate the development of PI-GNN mobility laws. We also expect the proposed framework to be beneficial when the dimensionality of the simulation parameters is even larger, e.g. for learning Non-Schmid effects in BCC. 

{Further work focusing on improving the PI-GNN mobility law using the proposed UQ-AL framework include (1) mixed modeling of the screw and mixed/edge dislocations, trained by MD data generated at different timescales for capturing both the fast-evolving mixed/edge dislocations and slow-evolving screw dislocations, (2) time scale-aware feature engineering for mobility law input, for enabling variable time-step integration, and (3) stochastic modeling of screw-like dislocations.} In addition to the mobility law, an important component of DDD simulations is the description of discrete events resulting in topological re-arrangements of the dislocation network. These consist of cross-slip of screw dislocations, junction formation and annihilation, and dislocation nucleation. Our PI-GNN mobility law is capable of predicting a 3-D displacement and thus, it is well set up to predict cross-slip events. Extension of the PI-GNN mobility law to describe dislocation junctions and nucleation is possible using advanced methodologies like dynamic graphs \cite{DBLP:journals/jmlr/KazemiGJKSFP20,DBLP:journals/tkde/ZhangWYLW23}, for example. We also note that the extension of our framework to more complex systems like alloys is possible by using additional features for local environment dependence. These advancements will be considered in future studies. 

We are aware of a pre-print by Bertin et al. \cite{bertin2023learning}, independently advocating a GNN-based method for learning dislocation mobilities. While the basic concept is similar, namely representing dislocations as graphs, there are multiple distinctions between the two approaches as follows: (1) Training protocols: we tested multiple, UQ-driven, training protocols, with relatively simple MD simulations of straight dislocations and expanding shear loops in broad parameter settings. In contrast, Bertin et al.~used large-scale MD simulations of high strain-rate deformation, using a narrow set of loading conditions; (2) Our PI-GNN approach allows the development of a physically interpretable non-linear mobility law, while such an interpretation is not straightforward in the approach by Bertin et al.; (3) Features for our PI-GNN model are composed of local stress tensor and line geometry (defined by Burgers vector and line tangent) as opposed to nodal forces in Bertin et al.; (4) Loss function in our approach is defined using a distance metric between nodal positions, as opposed to a Nye tensor based approach in Bertin et al. While there are pros and cons to the two approaches, a detailed evaluation of their relative merits will be the subject of a future study.

\section{Methods}
\label{sec:Methods}

In this section, we describe the generation of GT data from MD simulations, calculation of relevant features using the linear elasticity kernel in MoDELib Discrete Dislocation Dynamics (DDD) code, physics-informed graph neural networks (PI-GNN) framework for modeling mobility law of discrete dislocation dynamics, including the GNN architecture, physics-informed feature engineering, generalized loss function, and uncertainty-driven active learning (AL).

\subsection{Automated \textit{on-the-fly} generation of dislocation mobility data in MD}
\label{sec:MD}

Generation of dislocation motion data from molecular dynamics simulations has been widely addressed in past \cite{cereceda2012techniques,gilbert2011stress,maresca2018screw,queyreau2011edge}. It is a common practice to use free surface boundary conditions in the direction normal to the glide plane while building atomistic models with specific dislocation configurations moving under externally applied stress and temperature. Evaluation of the effect of free surfaces, on the stress experienced by the dislocation, requires solution of a boundary value problem coupled with the DDD framework.
To circumvent this, we used periodic boundary conditions and devised an automated model generation protocol where straight dislocation dipoles on a slip system $[11\bar{1}](101)$ are created for an arbitrary line orientation (defined by angle $\theta$) with the Burger's vector ($\pm \textbf{b}$). The simulation box is aligned as $b\sin {\theta}$ ($\textbf{x}$),  $b\cos {\theta}$ ($\textbf{y}$) and $[101]$  ($\textbf{z}$) with approximate lengths of 30 nm, 18 nm and 40 nm in these directions, adjusted based on the crystallography of Fe BCC unit cell. The dislocation line is oriented parallel to $\textbf{y}$ axis. Extra half-plane of atoms are removed along $\textbf{x}$ with a thickness of $b\sin {\theta}$. Note, for screw dipoles ($\theta=0^o$) this results in no deletion, while for an edge ($\theta=90^o$) a complete half-plane with a thickness of $b$ is removed. We use the periodic image correction superimposed on the Volterra displacement field with 4 periodic images on both sides along Cartesian directions $\textbf{x}$ and $\textbf{z}$ of the box, i.e., 63 [$(4\times2)^2 -1)$] images in total, following Cai et al \cite{cai2003periodic}. We report a few sample box-orientations in Table \ref{table:orientation details}, chosen out of a total 25 cases ranging from $0^o$ to $180^o$. 

For simulating the motion of a shear loop, an initial loop of radius $15$ nm is created on a $[11\bar{1}](101)$ system. The simulation box is then affinely deformed according to the linear elastic solution corresponding to the applied stress to prevent the loop from shrinking due to self interactions. To generate high-throughput dislocation trajectories, we subject both the loop and dipole configurations to a resolved shear stress range of $\sigma \in \{ 0.1, 0.25, 0.5, 0.75, 1.0\}$ GPa and a temperature range of $ T \in \{ 100, 300, 500, 700, 1000 \}$ K. All MD simulations were performed using the EAM potential for iron (Fe) developed by Mendelev et al.\cite{mendelev2003development}.

\begin{table}[htb]
\centering
\caption{List of few sample atomistic supercell orientations out of a total of 25 cases. Glide-plane normal $(101)$ is along $z$-axis for all the dipole configurations.} 
\begin{tabular}{c | c | c} 
 \hline
 Line angles w.r.t. Burger's vector ($\vec{b}$)  & $x$-axis orientation & $y$-axis orientation (line direction) \\ 
 \hline
 \\
 $0^o$ & $[1\bar{2}\bar{1}]$ & $[11\bar{1}]$\\ 
 $40.32^o$ & $[11\: \bar{4}\: \overline{11}]$ & $[2\,11\,\bar{2}]$\\
 $81.07^o$ & $[10\,7\,\overline{10}]$ & $[\bar{7}\,20\,7]$\\
 $110.41^o$ & $[28\,58\,\overline{28}]$ & $[\overline{29}\,28\,29]$\\
 $163.14^o$ & $[\bar{8}\,34\,8]$ & $[\overline{17}\,\bar{8}\,17]$\\ [1ex] 
 \hline
\end{tabular}
\label{table:orientation details}
\end{table}

To efficiently perform active learning, \textit{on-the-fly} processing of million atom MD trajectories is crucial. Our dipole models have 3-4 million atoms and MD simulations were performed using the kokkos-implementation of LAMMPS \cite{thompson2022lammps}. We apply a predefined global stress tensor and temperatures of the simulation box by running simulations in the NPT ensemble for 40 ps with a timestep of 1 fs. On NVIDIA A100 GPUs, 1 million EAM atoms/GPU can be simulated at a rate of up to 10ns/day, excluding any input/output (I/O) overhead.
To avoid reducing the simulation rate due to I/O as well as the storage of large volumes of MD data, we implement an \textit{in memory} and \textit{on-the-fly} analysis of dislocation trajectories via the Dislocation Extraction Algorithm (DXA) from Ovito-API \cite{stukowski2012automated}. Simulations were analyzed every 0.1 ps, with nodal spacings of $1.5$ nm. This is schematically shown in Figs. \ref{fig:Schematic}a and \ref{fig:Schematic}b.

\subsection{Augmenting coarse-grained MD data with local features}

Coarse-grained representations of dislocations present in the MD simulations are generated using the DXA algorithm. Using the same nodes identified by DXA and connecting the nodes via segments that are further discretized using Gauss points (GPs) (as shown in Fig.~\ref{fig:Schematic}b), an exact geometrical representation of the MD dislocation can be replicated in MoDELib, which is used to compute the local stress state at the GPs (Fig.~\ref{fig:Schematic}c). The stress state along the dislocation network is computed using a non-singular isotropic linear theory of dislocations \cite{cai2006non}, and it includes dislocation-dislocation interactions as well as the external stress  \cite{Lepinoux:1987tj,Ghoniem:1988wu,Gulluoglu:1989to,Kubin:1992wi,Schwarz:1997vo,Zbib:1998ub,Ghoniem:2000td,Weygand:2002tq,Bulatov:2004td,Po:2014en}.
MoDELib supports a periodic dislocation topology \cite{pachaury2022discrete}, and the stress field of the periodic dislocation network is computed using an Ewald summation strategy \cite{wells2015ewald}, which tapers the real-space long-range stress field by a user-defined characteristic length. Convergence of the stress calculation is typically realized using eight periodic images in each direction. 
Augmented coarse-grained MD snapshots are finally obtained, which consist of discretized dislocation lines using nodes, connectivities, and stress information at the GPs including both dislocation-dislocation interactions and external contributions.  
It is from these snapshots, containing both the geometry from MD and the local stress state computed by DDD that we move forward to construct the physical network and train the PI-GNN.  More details about the methodology can be found in Supplemental Information.

\subsection{Modeling mobility law using Physics-informed Graph Neural Network}
\label{sec:PIGNN}

\textit{Representing Dislocations using Heterogeneous graphs}. Graphs are a type of data structure that describes a set of objects, often referred to as vertices, and their relationships, or edges. Graphs are expressive and powerful at representing complex non-Euclidean data structures in social science, engineering, and physical science applications \cite{zhou2020graph,bronstein2021geometric}. In the context of dislocation dynamics, we demonstrate the mapping between the physical and computational representation of a dislocation network in Figs. \ref{fig:Schematic}d and \ref{fig:Schematic}e. Physical representation consists of nodes and GPs, connected by the dislocation segments, illustrated by Fig.~\ref{fig:Schematic}d. The physical features of the dislocation lines are defined on the GPs between two nodes, such as the Burger's vector, segment tangent vector, or local stress, while the features of the nodes consist only of positions. The data structure of a dislocation network can be easily mapped to a heterogeneous graph, where we introduce two types of vertices: $v_N$ representing nodes and $v_\text{GP}$ representing GPs in the physical network.

\begin{figure}[h]
    \centering
    \includegraphics[width=0.9\textwidth]{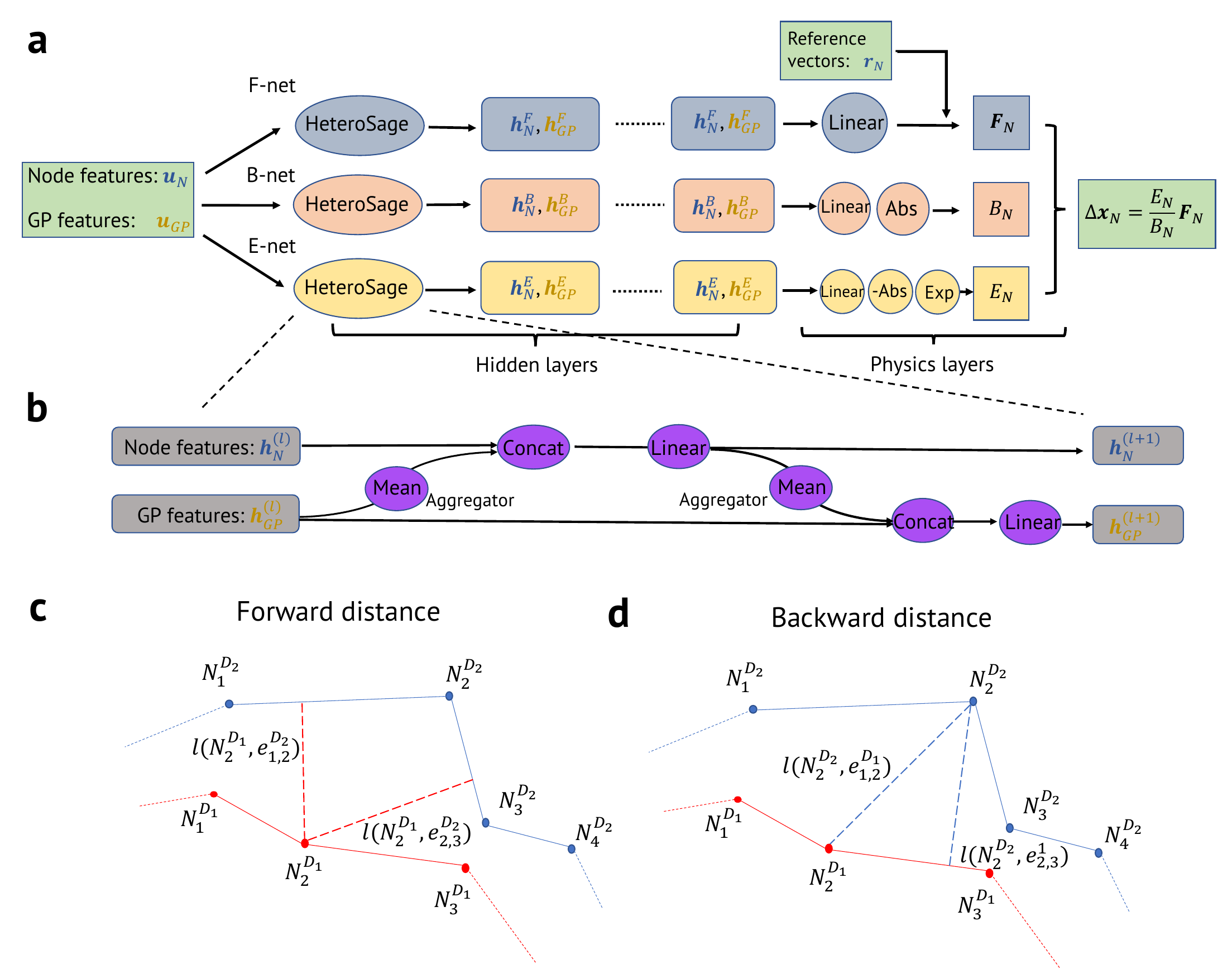}\caption{\textbf{Overview of the PI-GNN architecture and training loss function}. \textbf{a} Physics-informed graph neural network structure for learning the mobility law in Discrete Dislocation Dynamics (DDD), which consists of three different networks: F-net, B-net, and E-net. Each network consists of hidden layers, which are stacked HeteroSage GNNs, and physical layers, which are physically inspired functions adopted from a phenomenological description of dislocation mobility. \textbf{b} The proposed HeteroSage GNN structure and message passing diagram for heterogeneous graphs. \textbf{c,d} Illustrations of the proposed generalized loss function for measuring the distance between two dislocation configurations with different total number of nodes resulting from different discretization.}
    \label{fig:NN}
\end{figure}

\textit{Physics-informed graph neural network}. Figure \ref{fig:NN}a presents the structure of the graph neural network inspired by the physics-based phenomenological model of dislocation dynamics \cite{PO2016123} as shown in Eq.~\ref{eqn:DDDmodel}. We decompose the PI-GNN mobility law into three components: F-net, B-net, and E-net, each with distinct input features and network structures, depending on the physics-based mobility function. As shown in figure \ref{fig:NN}a, we model the three network components with two types of layers, hidden GNN layers, which consist of a series of black-box layers for learning feature embedding from the graph data structure, and physics layers, in which physical constraints for each component of the mobility function can be injected. With these two types of layers, we aim to combine the functional expressiveness of GNNs and the interpretability of physical structure to construct a powerful, robust, and predictive model for dislocation dynamics. For the hidden layers, we extend the widely used GraphSAGE \cite{hamilton2017inductive}, which were designed to operate on homogeneous graphs to sample and aggregate features from large graphs, to one that can operate on heterogeneous graphs. We refer to this extended model as HeteroSAGE in this work. Figure \ref{fig:NN}b illustrates the message-passing and feature operations of each HeteroSAGE layer: the features defined on the neighboring vertices are aggregated using an aggregator function, which is then concatenated with local features and passed through a fully-connected network. This process is then repeated in the opposite direction to generate the output of one HeteroSAGE layer. By stacking multiple HeteroSAGE layers, the model can learn complex feature embeddings from the graph. For the physics layers, we encode the mathematical structure of physics-based mobility law using differentiable functions. This design ensures that the output of each component adheres to the corresponding constraints dictated by the physics-based model. More details on the structure can be found in the Supporting Information.

\textit{Feature engineering}. In modern ML applications, deep neural networks, coupled with large datasets are commonly used to discover informative feature embeddings for effective predictions. However, in scientific applications, a set of carefully designed features can inject physics into the model, leading to improved prediction robustness and reduced dataset size requirements. In this work, we use a set of specially designed features from nodes and GPs as the input of the PI-GNN mobility law. We leverage translational invariance and rotational invariance, which are important in physical systems, for guiding the feature engineering processes. The input features of the PI-GNN are designed such that the aforementioned constraints are automatically satisfied by the model, thus injecting crucial physics-based induction biases into the model. Details on the selected features can be found in the Supplementary Sec.~\ref{sisec:featureengineering}.

\textit{Generalized loss function}. After constructing the GNN model, we define a loss function that quantifies the differences between the model-predicted evolution of the dislocation and the ground truth, whose minimization will constrain the parameters of the PI-GNN during learning. However, a challenge arises when formulating the loss function from coarse-grained MD simulation snapshots, as the total number of nodes in dislocations separated by the time step $\Delta t$ may differ and there is no clear one-to-one mapping between nodes. This hinders a straightforward computation of their differences using the commonly used Mean-Squared-Error (MSE) in traditional ML tasks. To resolve this issue, we propose a generalized loss function to measure the distances between two dislocations. As illustrated in figures \ref{fig:NN}c and \ref{fig:NN}d, the dislocations can be described as a collection of nodes and segments/edges connecting them. Instead of using the traditional point-wise positional difference to construct the loss function, we define a node-to-segment distance based on the geometric projection as:

\begin{eqnarray}
\label{eqn:ntos}
l(N_i^{D_1}, e_{j,k}^{D_2}) = \begin{cases} \l \vert {\bf x}'^{D_1}_{N_i}-{\bf x}^{D_1}_{N_i} \r\vert,{\bf x}'^{D_1}_{N_i} \in e^{D_2}_{j,k}, \\
\text{min}\l\{\l \vert {\bf x}^{D_1}_{N_i} -{\bf x}^{D_2}_{N_j} \r\vert, \l \vert {\bf x}^{D_1}_{N_i} -{\bf x}'^{D_2}_{N_{k}} \r\vert \r\}, {\bf x}'^{D_1}_{N_i} \notin e^{D_2}_{j,k}.
\end{cases}
\end{eqnarray}

With the node-to-segment distance $l(N_i^{D_1}, e_{j,k}^{D_2})$ defined, we use the minimum value of the node $N^{D_1}_i$ to all the segments $e^{D_2}_{j,k} \in D_2$ as the node-to-dislocation distance $l(N_i^{D_1}, D_2)$, which can then be used to compute the dislocation-to-dislocation distance as the sum of the node-to-dislocation distances $l(D_1, D_2) = \sqrt{\sum_{i} l(N_i^{D_1}, D_2)^2}$.
We notice that the proposed distance measure based on equations \ref{eqn:ntos}, is not symmetric, i.e. $l(D_1, D_2) \neq l(D_2, D_1)$. In this work, we adopt a loss function based on a symmetric distance measure,  which is achieved by averaging the distances computed in both directions to avoid the potential issue due to the smoothing effects (see Supplementary Sec.~\ref{sisec:generalloss} for more details):

\begin{eqnarray}
\label{eqn:loss}
L(D_1,D_2) &=& \frac{1}{2}\l[l(D_1, D_2) + l(D_2, D_1)\r], \\
Loss &=& \sum_{i} L(D^{\text{GT}}_i, D^{\text{Pred}}_i).
\end{eqnarray}

\subsection{Uncertainty quantification-drive active learning}
\label{sec:EnsembleLearning}
\textit{Ensemble learning for uncertainty quantification}. In the context of machine learning for scientific applications, particularly in cases where fully-resolved simulation data is sparse, the neural network-based model demonstrates competitive prediction accuracy. However, this may become problematic when a single, over-parameterized NN provides predictions for new datasets with unwarranted confidence. To address this issue, we adopt the ensemble learning approach to properly quantify the prediction uncertainty for modeling the mobility law using PI-GNN.

Ensemble learning is a machine learning training approach aimed at achieving improved prediction performance by combining the predictions from multiple independently-learned models. The objective of the ensemble learning approach lies in generating multiple instances of model parameters that span the local minima within the parameter-loss landscape of the NN-based model. We then approximate the PDF of the predictions using a Gaussian distribution, whose mean and variance are computed using the predictions from the top-performing ensemble models:
\begin{eqnarray}
\label{eqn:EnsemblePred}
\frac{\widehat{d\bf{x}}}{dt} &\approx& \overline{\mu}\left(\frac{d\bf{x}}{dt}\right) = \frac{2}{M_{en}} \sum_{i=1}^{M_{en}/2} \l. \frac{d\bf{x}}{dt} \r\vert_{\theta_i}, \\
\sigma^2\left(\frac{d{\bf x}_j}{dt}\right ) &\approx& \frac{2}{M_{en}} \sum_{i=1}^{M_{en}/2} \l (\l. \frac{d{\bf x}_j}{dt} \r \vert_{\theta_i}-\frac{\widehat{d{\bf x}}_j}{dt} \r)^2,
\end{eqnarray}
where ${d\bf{x}}/{dt}\vert_{\theta_i}$ refer to the PI-GNN prediction of the nodal displacement using the parameter $\theta_i$. Through the ensemble learning procedure, we can obtain an estimate of the distribution of PI-GNN outputs, which can inform the UQ of predictions.

\label{sec:ActiveLearning}
\textit{Active learning}. Active learning is a special procedure in ML that can interactively query a new dataset based on a pre-determined strategy. The active learning approach is commonly used in supervised learning setups where it is very expensive to manually enumerate new data points for the query. When a proper strategy for selecting new data points is used, the total number of data points to learn a model can be dramatically reduced compared to a passive learning strategy with a randomly queried dataset.

Figure \ref{fig:MDALLoopsummary} shows the workflow for the UQ-based active learning loop, which includes data generation, model training, UQ and data point selection. The details of the active learning procedure are described below:

\begin{enumerate}
    \item Generate initial MD dataset with a set of randomly selected simulation input features $\Lambda = \l \{ \lambda_i\r \}$, where $\lambda_i$ can be different combinations of temperature, applied stress, orientations, Burger's vector, etc. The simulated MD trajectories are then coarse-grained and post-processed to obtain snapshots of the dislocations nodes/GPs and their features. We denote the MD dataset with parameters $\Lambda$ as ${\bf D}(\Lambda)$.
    \item Train an ensemble of $M_{en}$ PI-GNNs on the dataset ${\bf D}(\Lambda)$ each with random initialization of model parameters. We employ a train/test/validation data split ratio of $70\%/20\%/10\%$.
    \item Randomly sample a new set of MD simulation input features $\Lambda'$ and run new MD simulations with them and regroup with existing test dataset to generate a test dataset ${\bf D}_t({\Lambda'})$.
    \item Take the top $50\%$ ($M_{en}/2$) performing trained PI-GNN models and use the ensemble variance of the nodal displacement prediction as a measure of prediction uncertainty.
    \item If the highest uncertainty level is lower than a preset threshold, stop training. Otherwise, combine the most uncertain set of the test data with the training data to form a new set of training data, i.e. ${\bf D}(\Lambda \cup \Lambda') $, which is then used to re-train (using previously converged model parameters as the initial weights) the ensemble models. 
    \item Move to step 3.

\end{enumerate}

\section*{Acknowledgements} \label{sec:acknowledgements}
The team is supported by the Laboratory Directed Research and Development (LDRD) Project ``Accelerated Dynamics Across Computational and Physical Scales'' (20220063DR). The study has been approved for unlimited release with LA-UR-23-26562. The authors acknowledge significant support from the Darwin test bed at Los Alamos National Laboratory (LANL), which is funded by the Computational Systems and Software Environments subprogram of LANL's Advanced Simulation and Computing program (NNSA/DOE), and Institutional High Performance Computing.
\bibliographystyle{unsrt}
\bibliography{refs_arxiv}

\end{document}


\title{Supplementary Information for ``Data-Driven Modeling of
Dislocation Mobility from Atomistics using Physics-Informed Machine Learning''}


\author[1]{Yifeng Tian}

\author[2]{Soumendu Bagchi}

\author[3]{Liam Myhill}

\author[4]{Giacomo Po}

\author[3]{Enrique Martinez}

\author[1]{Yen Ting Lin}

\author[2]{Nithin Mathew}

\author[2]{Danny Perez}

\affil[1]{Information Sciences Group, Computer, Computational and Statistical Sciences Division (CCS-3), Los Alamos National Laboratory, Los Alamos, 87545, NM, USA}

\affil[2]{Physics and Chemistry of Materials , Theoretical Division (T-1), Los Alamos National Laboratory, Los Alamos, 87545, NM, USA}

\affil[3]{Department of Materials Science and Engineering, Clemson University, Clemson, 29623, SC, USA}

\affil[4]{Department of Mechanical and Aerospace Engineering, University of Miami, Miami, 33146, FL, USA}
\date{}

\maketitle

\makeatletter\@input{xx.tex}\makeatother
\section{Methods}

 Additional details to describe the calculation of relevant physical quantities using the linear elasticity kernel in Modelib Discrete Dislocation Dynamics (DDD) code, Physics-informed Machine Learning based on Graph Neural Networks (PI-GNN) framework for modeling discrete dislocation dynamics, including the GNN architecture, physics-informed feature engineering, generalized loss function, and uncertainty-driven active learning (AL) are provided here.

\subsection{Computing the stress tensor at dislocation segments}
\label{sisec:computstress}

A point to note is that  DXA does not inherently provide dislocation snapshots as  closed-loops, which is one of the constraining features of the MoDELib DDD framework. The nodes populated by DXA may contain duplicate information, which must be removed before it can be used by ModeLib for stress computation. This is critical, as the density of nodes near the quadrature point  to will affect the computation of the dislocation self-stress. 
Therefore, a method for filtering and cleaning the DXA analysis was developed and generalized to work for complex dislocation networks containing a multitude of dislocations with varied defining features. Once cleaned, the data can be passed into ModeLib seamlessly and used to train the PI-GNN.
\subsection{Modeling mobility law using Physics-informed Graph Neural Network (PI-GNN)}

\subsubsection{Heterogeneous graph}

Graphs are a type of data structure that describes a set of objects, often referred to as nodes or vertices, and their relationships, or edges. Graphs are expressive and powerful at representing complex non-Euclidean data structures in social science, engineering, and physical science applications \cite{zhou2020graph,bronstein2021geometric}. Dislocation networks, represented as a set of nodes and Gauss points in DDD, can be naturally encoded as a graph.  We will use the Graph Neural Network (GNN), a neural architecture designed to operate on graphs, for directly learning dislocation mobility from coarse-grained Molecular Dynamics (MD) simulation data. 

Here, we denote a graph as $\bf{G}=(V,E)$, where $\bf{V}$ is the collection of nodes or vertices and $\bf{E}$ is the set of edges. To avoid confusion, we will refer to a node/vertex in a graph as a vertex, and a node in the dislocation dynamics configuration as a node.  For a vertex $v \in {\bf V}$, we denote the set of its neighboring vertices as $\mathcal{N}(v)$. For any $v' \in \mathcal{N}(v)$, the edge $(v,v') \in {\bf E}$ connects two vertices $v$ and $v'$. 

In the context of dislocation dynamics, we can represent the dislocation nodes and GPs as the vertices $v \in \bf{V}$ in a graph, and the dislocation segments that define the connection between nodes and GPs as edges $ e \in \bf{E}$. The features of a dislocation node $v$ can be represented as ${\bf u}_v$ and it may include information that describes the state of the dislocation node. A challenge arises when we attempt to represent a dislocation network as a homogeneous graph, which requires that all the nodes in a graph contain the same type of features. However, in the dislocation dynamics, the nodes and GPs have distinctly different types of features. Dislocation nodes contain only the positional information. The GPs contain the information of Burgers vector, stress, and line orientations. As such, it is not possible to represent a dislocation network by a homogeneous graph, where the features of all the vertices have the same dimension. To resolve this issue, we introduce the heterogeneous graph, which can be used to describe graph data structures with different types of vertices.

\begin{figure}[htb] 
    \centering
    \includegraphics[width=0.8\textwidth]{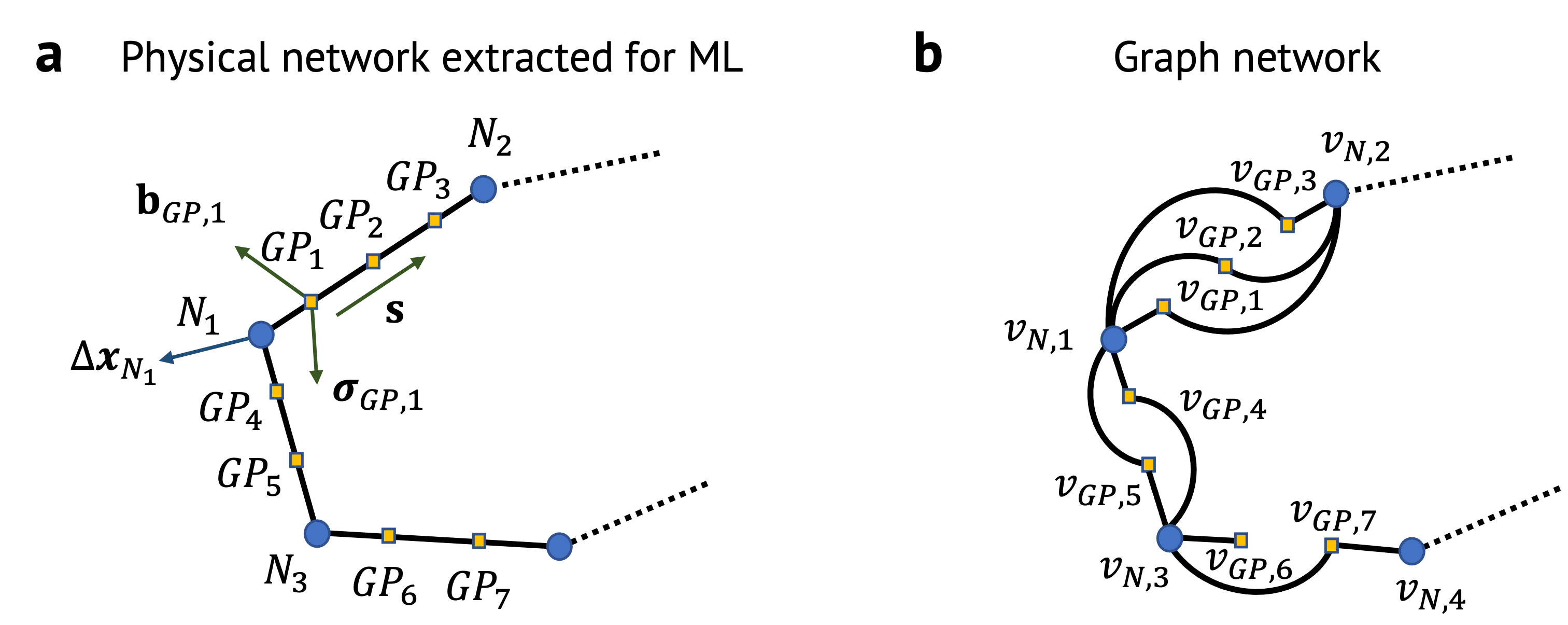}
    \caption{\textbf{Schematic of the representations of the dislocation netowrk:} \textit{(a)} General representation of the dislocation configuration with multiple Gauss points between nodes. \textit{(b)} Graph representation of the dislocation configuration.}
    \label{sifig:Schematic}
\end{figure}

In Figure \ref{sifig:Schematic} (a) and (b), we demonstrate the mapping between the physical and computational representation of a dislocation network. Physical representation consists of nodes and GPs, connected by the dislocation segments, illustrated by Figure \ref{sifig:Schematic} (a). The physical features of the dislocation lines are defined on the GPs between two nodes, such as the Burger's vector ${\bf b}$, segment tangent vector ${\bf s}$, and local stress ${\bf \sigma}$, while the features of the nodes consist only of positions. Computationally, the GPs are used as quadrature points for constructing local stresses using a constitutive model that provides physics-based features for modeling dislocation dynamics. Figure \ref{sifig:Schematic} (b) illustrates the corresponding heterogeneous graph we will use for performing learning. We introduce two types of vertices in the heterogeneous graph, $v_\text{N}$ and $v_\text{GP}$, which represent the nodes and GPs in the physical network respectively. The feature space of the two types of vertices is different, which will be discussed in more detail in the next section. In the graph representation, instead of connecting GPs $v_\text{GP}$ sequentially into a segment between nodes $v_\text{N}$, we directly connect  GPs to their nearest nodes, so that the information on GPs can be passed directly to the nearby nodes (and vice versa). 

\begin{figure}[h]
    \centering
    \includegraphics[width=0.8\textwidth]{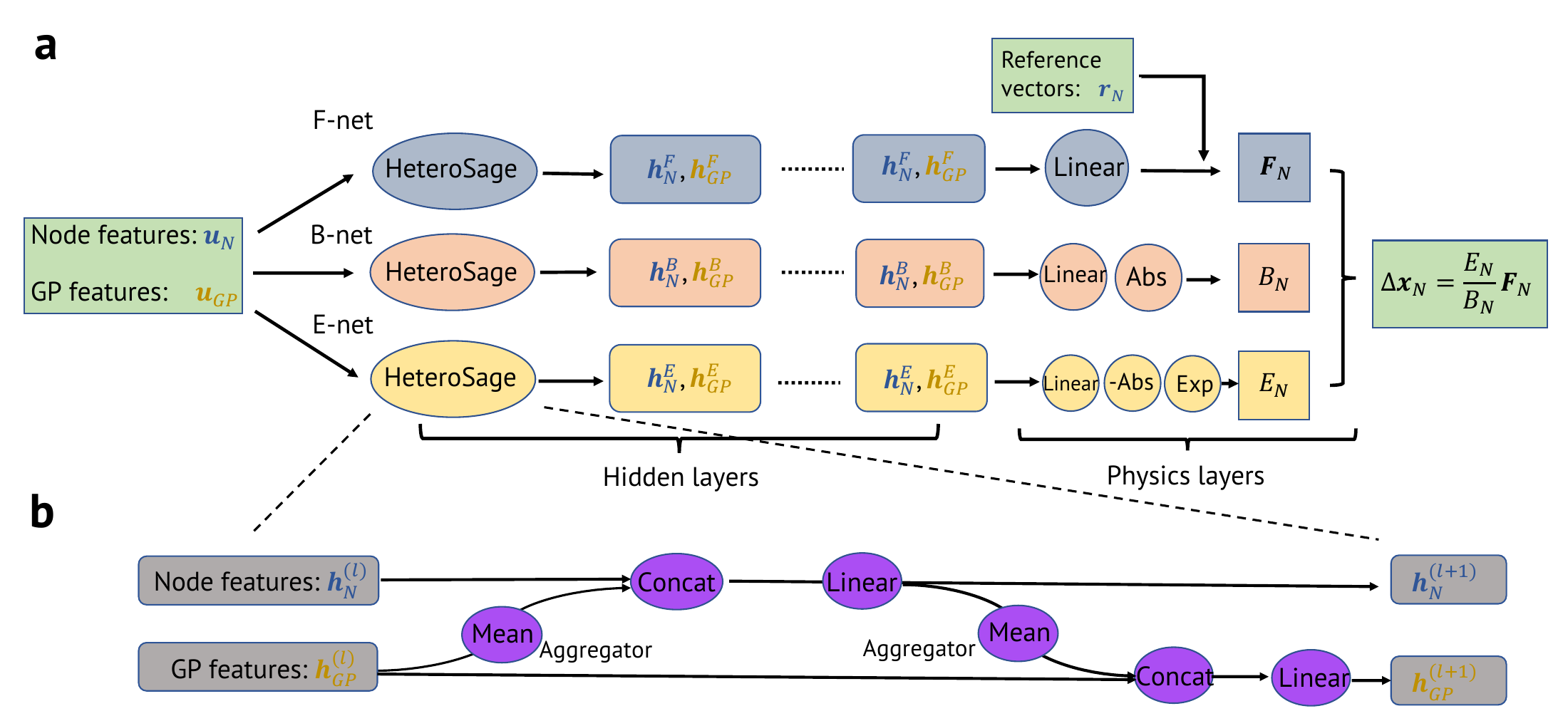}
    \caption{(\textit{a}) Physics-informed Graph Neural Network structure for learning the mobility law in Discrete Dislocation Dynamics (DDD), which consists of three different networks: F-net, B-net, and E-net. Each network consists of hidden layers, which are stacked HeteroSage GNNs, and physical layers, which are physically inspired functions adopted from a phenomenological description of dislocation mobility. (\textit{b}) The proposed HeteroSage GNN structure and message passing diagram for heterogeneous graphs.}
    \label{fig:NN1}
\end{figure}

\subsubsection{Physics-informed Graph Neural Networks}
\label{sisec:PI-GNN}
Graph neural networks take features on graph vertices and graph edges as input and learn a mapping to the target output by message-passing \cite{bronstein2021geometric}. In this work, we aim to use the GNN to learn the mobility law $\mathbf{M}$, which characterizes the stress- and temperature-dependence of the instantaneous velocity of the dislocation lines.

The proposed physics-informed mobility law maps from the configuration of dislocation line $D_0$ at time $t_0$ to its future configuration $D_1$ at $t_1= t_0+\Delta t$, $D_1 = M(G(D_0), {\bf u}^{D_0}_N, {\bf u}^{D_0}_{GP})$, where $G(D_0)$ is the heterogeneous graph built on dislocation configuration $D_0$, and ${\bf u}_N^{D_0}$, ${\bf u}_{GP}^{D_0}$ represent the nodes and GPs' features on dislocation $D_0$. The proposed PI-GNN mobility law is no longer a function of local features like in the traditional mobility law. Instead, it leverages GNN and graph data structures to learn both local/non-local dependencies on the features and configurations that have not been considered in the traditional mobility law.

In order to operate on the Heterogeneous graph data structure shown in figure \ref{fig:NN1} (b), we propose an extension of the widely used GraphSAGE \cite{hamilton2017inductive}, which was designed to operate on homogeneous graphs to sample and aggregate features from large graphs to one that can operate on heterogeneous graph. We refer to the proposed GNN model as HeteroSAGE in this paper.

In GraphSAGE, an aggregator function, which can either be parametrized/learned or prescribed, is first used to aggregate feature information from a vertex's local neighborhoods. The aggregated information from all the neighboring vertices is then grouped with the local information to generate a set of feature embeddings using an NN. Then these aggregator functions and NNs are layered into a deep ML model so that the GraphSAGE is capable of aggregating and learning complex features from different numbers of hops away from any specified vertex.

Here, we modify and adapt the structure of GraphSAGE for operating on heterogeneous graphs. For $i$-th layer of the proposed HeteroSAGE, the forward operation on the heterogeneous graph constructed by nodes and GPs can be expressed as:

\begin{eqnarray}
\label{eqn:sage}
{\bf m}_{N,i}^{(l+1)} &=& \text{Aggregate}\l(\l \{{\bf h}_{GP,j}^{(l)}. \forall v_{GP,j} \in \mathcal{N}(v_{N,i})\r\}\r), \\
{\bf h}_{N,i}^{(l+1)} &=& \text{ReLU}\l ( {\bf W}^{(l)}_N \cdot \text{Concat} \l({\bf h}_{N,i}^{(l)}, {\bf m}_{N,i}^{(l+1)} \r) +{\bf b}^{(l)}_N \r ), \\
{\bf m}_{GP,i}^{(l+1)} &=& \text{Aggregate}\l(\l \{{\bf h}_{N,j}^{(l)}. \forall v_{N,j} \in \mathcal{N}(v_{GP,i})\r\}\r), \\
{\bf h}_{GP,i}^{(l+1)} &=& \text{ReLU}\l ( {\bf W}^{(l)}_{GP} \cdot \text{Concat} \l({\bf h}_{GP,i}^{(l)}, {\bf m}_{GP,i}^{(l+1)} \r) +{\bf b}^{(l)}_{GP}\r ).
\end{eqnarray}

For each HeteroSAGE layer, a heterogeneous graph ${\bf G}=({\bf V}, {\bf E}) $ that consists of two types of vertices, and their corresponding features, ${\bf h}_N^{(l)}$ and $ {\bf h}_{GP}^{(l)}$ needs to be provided as the inputs. Figure \ref{fig:NN1} (b) demonstrates the steps of operations in each layer, which can be summarized into the following steps. At $l$-th layer of the GNN, each node vertex $v_{N}$ aggregates the features from its immediate neighbors of GP vertices $\{ {\bf h}^{(l)}_{GP}, \forall v_\text{GP}\in \mathcal{N}(v_{N})\}$. Note that the aggregation step is operated on the hidden features/embeddings generated from the previous layer and when $l=1$, the hidden features are the input node features. The aggregator function is defined by the user and it can either be a fixed or trainable function. In this work, we use a Mean aggregator function to take element-wise mean of the neighboring features $\{ {\bf h}^{(l)}_{GP}, \forall v_\text{GP}\in \mathcal{N}(v_{N})\}$, which is then stored in the message variable in ${\bf m}^{(l+1)}_{N}$. After aggregating information from neighboring vertices, the HeteroSAGE then concatenate current hidden features vector ${\bf h}_N^{(l)}$ and the aggregated neighboring features vector ${\bf m}^{(l+1)}_{N}$. This concatenated features vector is then fed into a fully-connected NN that operate on node vertices with weight ${\bf W}_N^{(l)}$ and bias ${\bf b}_N^{(l)}$ and then a ReLU activation function $\sigma$, whose output becomes the embedded feature of node for the input of next layer. The second part of the HeteroSAGE is similar to the previous operation, but it passes message in the opposite direction, from nodes to GPs. In the aggregation step, the hidden features of the node vertices, which are the immediate neighbors of a given GP vertex $v_\text{GP}$, $\{{\bf h}^{(l)}_{N}, \forall v_{N} \in \mathcal{N}(v_\text{GP})\}$ are fed into the same aggregator function to form a single feature vector ${\bf m}_{GP,i}^{(l+1)}$. It is then concatenated with the hidden features of GP from last layer can fed into another fully-connected NN with weight ${\bf W}_{GP}^l$, bias ${\bf b}^l_{GP}$ , and activation function to generate the hidden features for the input of next layer. With this adaptation of the GraphSAGE, we are able to apply it on heterogeneous graph with vertices that feature vectors of different dimensions.

After we define the most important component of the GNN model for modeling dislocation dynamics, we can then use it to construct the physics-informed GNN model. Figure \ref{fig:NN1}(a) shows the structure of the PI-GNN model.

We follow the mathematical structure of the phenomenological model of dislocation dynamics mobility law \cite{PO2016123}:

\begin{eqnarray}
\label{sieqn:DDDmodel}
\frac{d{\bf x}}{dt} = \frac{{\bf f}( {\bf \sigma}, {\bf b}, {\bf s})}{B(T, {\bf b}, {\bf s})} exp\l (-\frac{\epsilon({\bf \sigma}, {\bf b}, {\bf s},T)}{k_B T}\r),
\end{eqnarray}

where ${\bf f}$, $B$, and $\epsilon$ are functions representative of the force, drag, and activation energy in dislocation dynamics. The inputs for the physics-based mobility function, ${\bf \sigma}$, ${\bf b}$ and ${\bf s}$ are local stress tensor, Burger's vector and segment tangent vector. We embed this prior knowledge based on physical intuition into the construction of the PIML GNN model and decompose the ML modeling into three components: F-net, B-net, and E-net. The mobility law represented by the PI-GNN is then:

\begin{eqnarray}
\label{eqn:DDDmodelGNN}
\frac{d {\bf x}_N}{dt} = {\bf M}(G, {\bf u}_{N}, {\bf u}_{GP}) = \frac{E_{GNN}(G, {\bf u}_{N}^{E}, {\bf u}_{GP}^{E})}{B_{GNN}(G, {\bf u}_{N}^{B}, {\bf u}_{GP}^{B} )}  {\bf F}_{GNN}(G, {\bf u}_{N}^{F}, {\bf u}_{GP}^{F}),
\end{eqnarray}

Note that the input features for the three networks are different, depending on the physics-based mobility function in Eq.~\ref{sieqn:DDDmodel}. As shown in figure \ref{fig:NN1} (a), we model the three network components with two types of layers, Hidden GNN layers, which consist of a series of black-box HeteroSAGE layers for learning feature embedding from the graph structure, and physics layers, in which hard physical constraints for each component of the mobility function can be injected. With these two types of layers, we intend to combine the functional expressiveness of GNN with over-parametrization and interpretability of physical structure to construct a powerful, robust, and predictive model for dislocation dynamics.

\subsubsection{Feature engineering}
\label{sisec:featureengineering}
In modern ML applications, a deep neural network with a large data set is often used to discover embedded features that is informative for making predictions. However, in scientific applications, a carefully designed set of features can help inject physics into the model, improving the robustness of the ML model prediction, and reducing the required size of the data set. In this work, we use a set of specially designed features from nodes and GPs as the input of the PI-GNN.
 
In discrete dislocation dynamics, the following properties of the nodes and GPs are used to predict the node displacement using mobility function: positions ${\bf x}_N$ defined at nodes, positions ${\bf x}_{GP}$, burgers vector ${\bf b}_{GP}$, local stress ${\bf \sigma}_{GP}$, and segment tangent ${\bf s}_{GP}$ defined at GPs and temperature ($T$) of the system. Here we consider physical invariance, such as translational invariance and rotational invariance as physical constraints to design features. Let us consider the operator that represents the physical transformations as $\mathcal{T}$, which can be any composition of the Galilean, translational, and rotation transformation. We require that the designed features and the class of functions represented by the PI-GNN is $\mathcal{T}$-invariant:

\begin{equation}
    \mathcal{T}{\bf M}(G, {\bf u}_N, {\bf u}_{GP}) = {\bf M}(G, \mathcal{T}{\bf u}_N, \mathcal{T}{\bf u}_{GP}).
\end{equation}

When the transformation operator $\mathcal{T}$ represents spatial translation, i.e. $\mathcal{T}{\bf x} = {\bf x} +{\bf x}'$, a straightforward approach to ensure the invariance of the input features is to use the relative displacement vector of the nodal and GPs' positions instead of their absolute positions. Other properties, such as the Burger's vector, local stress and temperature are not affected by the translational operation. For rotational invariance/equivariance, we need to ensure the input features are rotational invariant and vector outputs $\frac{d {\bf x}}{dt}$ are equivariant. For the input features, we decompose the vector/tensor properties into its magnitude/eigenvalues and tangent/eigenvectors. The magnitude/eigenvalues are scalars that are already rotational invariant, which will be used as inputs for the PI-GNN mobility function. For tangent/eigenvectors, we use their relative angles to a locally defined frame of reference based on the segments vector ${\bf s}$ and glide plane normal ${\bf n}$ to convert them to a set of scalar input features. At the output layer, we decompose the vector output $\frac{d {\bf x}}{dt}$, or more specifically the F-net ${\bf f}_N$, into a linear combination of the equivariant vector bases (similarly defined using the local frame of reference) and scalar functions of the invariant scalar inputs ${\bf f}_N = \sum_{i=1}^3 f_i(G, {\bf u}_N, {\bf u}_{GP})  {\bf r}_i$, where $f_i$'s are the outputs of the F-net and ${\bf r}_i$ are locally defined bases vectors.

To summarize, the input features of the nodes and GPs for the three GNN components are:

\begin{eqnarray}
\text{F-net Node features:}&& \quad \l .\overline{\l \vert {\bf s}_{ij}\r \vert} \r \vert_{N_j \in \mathcal{N}(N_{i})}  , \quad  \l . \overline{ .\frac{{\bf s}_{ij} \cdot {\bf s}_{ik}}{\vert {\bf s}_{ij}\vert \vert{\bf s}_{ik} \vert}} \r\vert_{N_{j}, N_{k} \in \mathcal{N}(N_{i}), j\neq k}  , \\
\text{F-net GP features:}&& \quad \frac{{\bf b} \cdot {\bf s}}{\vert {\bf s}\vert}, \quad e_1, \quad e_2, \quad e_3, \quad \frac{{\bf v}_1\cdot{\bf s}}{\vert {\bf s} \vert}, \quad \frac{{\bf v}_2\cdot{\bf s}}{\vert {\bf s} \vert}, \quad \frac{{\bf v}_3\cdot{\bf s}}{\vert {\bf s} \vert}, \quad \\
\text{B-net Node features:} && \quad \l .\overline{\l \vert {\bf s}_{ij}\r \vert} \r \vert_{N_j \in \mathcal{N}(N_i)}  , \quad  \l . \overline{ \frac{{\bf s}_{ij} \cdot {\bf s}_{ik}}{\vert {\bf s}_{ij}\vert \vert{\bf s}_{ik} \vert}} \r\vert_{N_j, N_k \in \mathcal{N}(N_i), j\neq k}  , \\ 
\text{B-net GP features:}&& \quad \frac{{\bf b} \cdot {\bf s}}{\vert {\bf s}\vert}, \quad T, \\
\text{E-net Node features:} && \quad \l .\overline{\l \vert {\bf s}_{ij}\r \vert} \r \vert_{N_j \in \mathcal{N}(N_i)}  , \quad  \l . \overline{ \frac{{\bf s}_{ij} \cdot {\bf s}_{ik}}{\vert {\bf s}_{ij}\vert \vert{\bf s}_{ik} \vert}} \r\vert_{N_j, N_k \in \mathcal{N}(N_i), j\neq k} , \\ 
\text{E-net GP features:}&& \quad \frac{{\bf b} \cdot {\bf s}}{\vert {\bf s}\vert}, \quad e_1, \quad e_2, \quad e_3, \quad \frac{{\bf v}_1\cdot{\bf s}}{\vert {\bf s} \vert}, \quad \frac{{\bf v}_2\cdot{\bf s}}{\vert {\bf s} \vert}, \quad \frac{{\bf v}_3\cdot{\bf s}}{\vert {\bf s} \vert} \quad, T
\end{eqnarray}

where ${\bf s}_{ij}$ are the relative distance vector from node $N_i$ to $N_j$. $e_i$'s and ${\bf v}_i$'s are the eigenvalues and eigenvectors of the local stress. The bases vector for the local reference of frame ${\bf r}_i$ is defined as:

\begin{eqnarray}
{\bf r}_1 = \l . \frac{\overline{ {\bf s}_{ij}} }{\l \vert\overline{ {\bf s}_{ij}}\r \vert} \r\vert_{N_j \in \mathcal{N}(N_i)}, \quad {\bf r}_2 = {\bf n}, \quad {\bf r}_3 = {\bf r}_1 \times {\bf r}_2,
\end{eqnarray}
where ${\bf n}$ denotes the glide plane normal and $\times$ denote vector cross product.

\subsection{Generalized loss function}
\label{sisec:generalloss}
After constructing the GNN model, we need to define a loss function that measures the differences between the model predicted evolution of the dislocation and the ground truth, which are nodal positions of dislocations, as identified by DXA, from the MD simulations after a certain time step $\Delta t$. A challenge for defining the loss function from coarse-grained MD simulation snapshots is that the total number of nodes in dislocations separated by the time step $\Delta t$ may differ and there is no clear one-to-one mapping for straightforward computation of their differences, like Mean-Squared-Error (MSE) commonly used in traditional ML tasks. The recently proposed Sweep-Tracing Algorithm \cite{BERTINSweep} provides a computational algorithm to connect pairs of successive DXA snapshots. However, the simpler algorithm proposed below was used in our simulations.

\begin{figure}[h]
    \centering
    \includegraphics[width=0.8\textwidth]{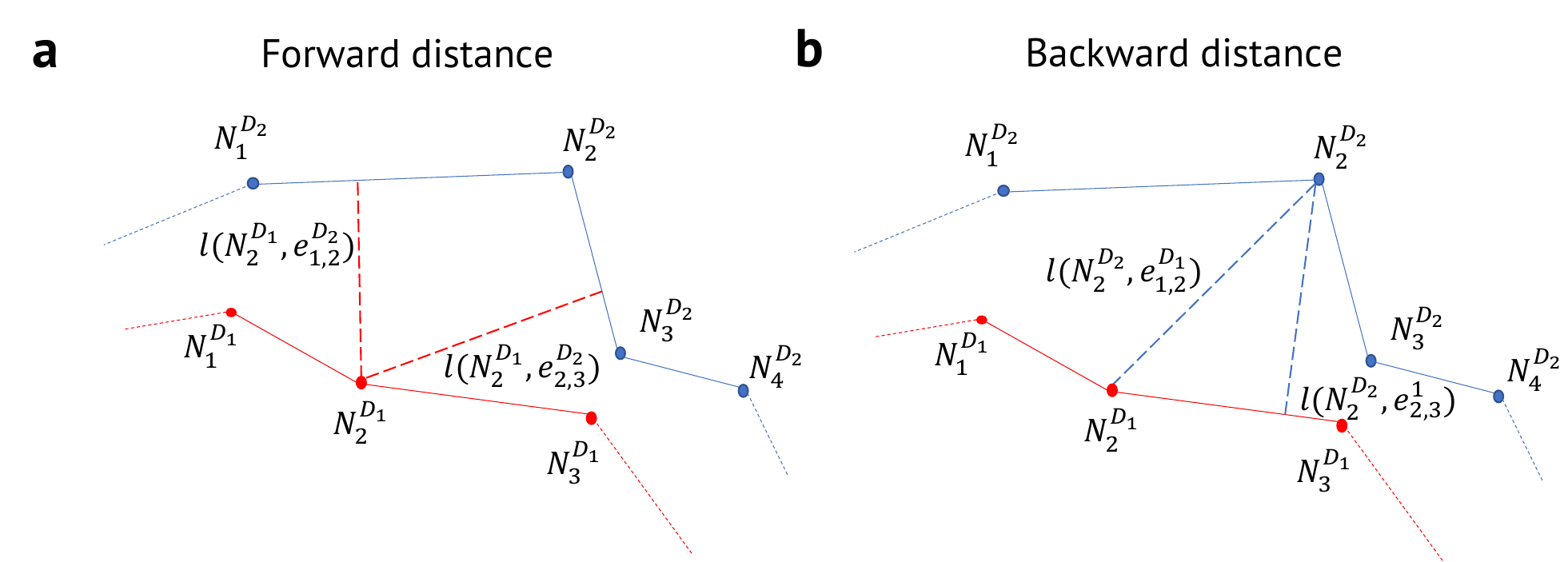}
    \caption{(\textit{a, b}) Illustrations of the proposed generalized loss function for measuring the distance between two dislocation configurations with different numbers of nodes.}
    \label{fig:NN2}
\end{figure}

To resolve this issue, we propose a generalized loss function to measure the difference between two dislocations. As illustrated in figures \ref{fig:NN2} (a) and (b), the dislocations can be described as a collection of nodes and segments/edges between them. Here we use $D_1$ and $D_2$ to denote the two dislocations. Note here, we do not include the GPs in the descriptions of the dislocation, because GPs are used for computing local stress in DDD, and do not provide additional information for the dislocation configuration. For two dislocations $D_1$ and $D_2$, we denote their nodes as $N_i^{D_1}$, $N_j^{D_2}$, their positions as ${\bf x}^{D_1}_{N_{i}}$, ${\bf x}^{D_2}_{N_j}$, and the segments that connect two nodes $N_j$ and $N_{k}$ in dislocation $D_1$ as $ e^{D_1}_{j,k}$. Note the difference between ${\bf s}_{j,k}$, which is the segment tangent vector, and $e_{j,k}$, which represent the collection of points between ${\bf x}_{N_{j}}$, ${\bf x}_{N_k}$. To construct the generalized loss function, we start by measuring the distance from a given node $N_i^{D_1}$ to a segment ${\bf s}^{D_2}_{j,k}$ between two nodes $N_j^{D_2}$, $N_k^{D_2}$. Here we use $l$ to denote the distance between two objects, which could be nodes, segments, and dislocations. We first compute the node to segment distance $l(N_i^{D_1},  e_{j,k}^{D_2} )$ by projecting ${\bf x}^{D_1}_{N_i}$ onto the segment $({\bf x}^{D_2}_{N_j}, {\bf x}^{D_2}_{N_{k}})$ to obtain the projection point ${\bf x}'^{D_1}_{N_i}$. Then this node-to-segment distance $l(N_i^{D_1},  e_{j,k}^{D_2} )$ can be calculated using the following equation:

\begin{eqnarray}
\label{sieqn:ntos}
l(N_i^{D_1}, e_{j,k}^{D_2}) = \begin{cases} \l \vert {\bf x}'^{D_1}_{N_i}-{\bf x}^{D_1}_{N_i} \r\vert,{\bf x}'^{D_1}_{N_i} \in e^{D_2}_{j,k}, \\
\text{min}\l\{\l \vert {\bf x}^{D_1}_{N_i} -{\bf x}^{D_2}_{N_j} \r\vert, \l \vert {\bf x}^{D_1}_{N_i} -{\bf x}'^{D_2}_{N_{k}} \r\vert \r\}, {\bf x}'^{D_1}_{N_i} \notin e^{D_2}_{j,k}.
\end{cases}
\end{eqnarray}

With the node-to-segment distance $l(N_i^{D_1}, e_{j,k}^{D_2})$ defined, we use the minimum value of the node $N^{D_1}_i$ to all the segments $e^{D_2}_{j,k} \in D_2$ as the node-to-dislocation distance $l(N_i^{D_1}, D_2)$. 

\begin{eqnarray}
\label{eqn:ntod}
l(N_i^{D_1}, D_2) = \text{min}\l\{l(N_i^{D_1}, e_{j,k}^{D_2}),  \forall e_{j,k}^{D_2} \in D_2 \r\}
\end{eqnarray}
Then the distance between two dislocations can be computed using the sum of all node-to-dislocation distances:

\begin{eqnarray}
\label{eqn:dtod}
l(D_1, D_2) = \sqrt{\sum_{i} l(N_i^{D_1}, D_2)^2} .
\end{eqnarray}

We notice that the proposed distance measure based on equations \ref{sieqn:ntos}, \ref{eqn:ntod} and \ref{eqn:dtod} is not symmetric, i.e. $l(D_1, D_2) \neq l(D_2, D_1)$. By using only the one-side distance measure, we noticed through numerical experiment that the optimization tends to smooth out the dislocation line. Therefore in this work, we adopt a loss function based on a symmetric distance measure, which is achieved by averaging the distance in equation \ref{eqn:dtod} in both directions:

\begin{eqnarray}
\label{sieqn:loss}
L(D_1,D_2) &=& \frac{1}{2}\l[l(D_1, D_2) + l(D_2, D_1)\r], \\
Loss &=& \sum_{i} L(D^{\text{GT}}_i, D^{\text{Pred}}_i),
\end{eqnarray}

where $D^{\text{Pred}}_i$ is $i$-th predicted configuration of the dislocation, whose nodal positions are computed as:
\begin{eqnarray}
{\bf x}_{N}^{\text{Pred}} = {\bf x}_{N}^{\text{Init}} + \Delta {\bf x}_N, \\
\Delta {\bf x}_N  = \frac{d {\bf x}_N}{dt} \Delta t.
\end{eqnarray} 

 We have tested that with the proposed generalized loss function for discretized dislocations lines, we can properly learn a function that describes their differences.

\section{Additional results and discussions}

\subsection{Additional results on the prediction quality of the learned PI-GNN mobility law}
\label{sisec:pignnresults}

\begin{figure}[htb]
    \centering
    \includegraphics[trim=40 100 40 80, width=1\textwidth]{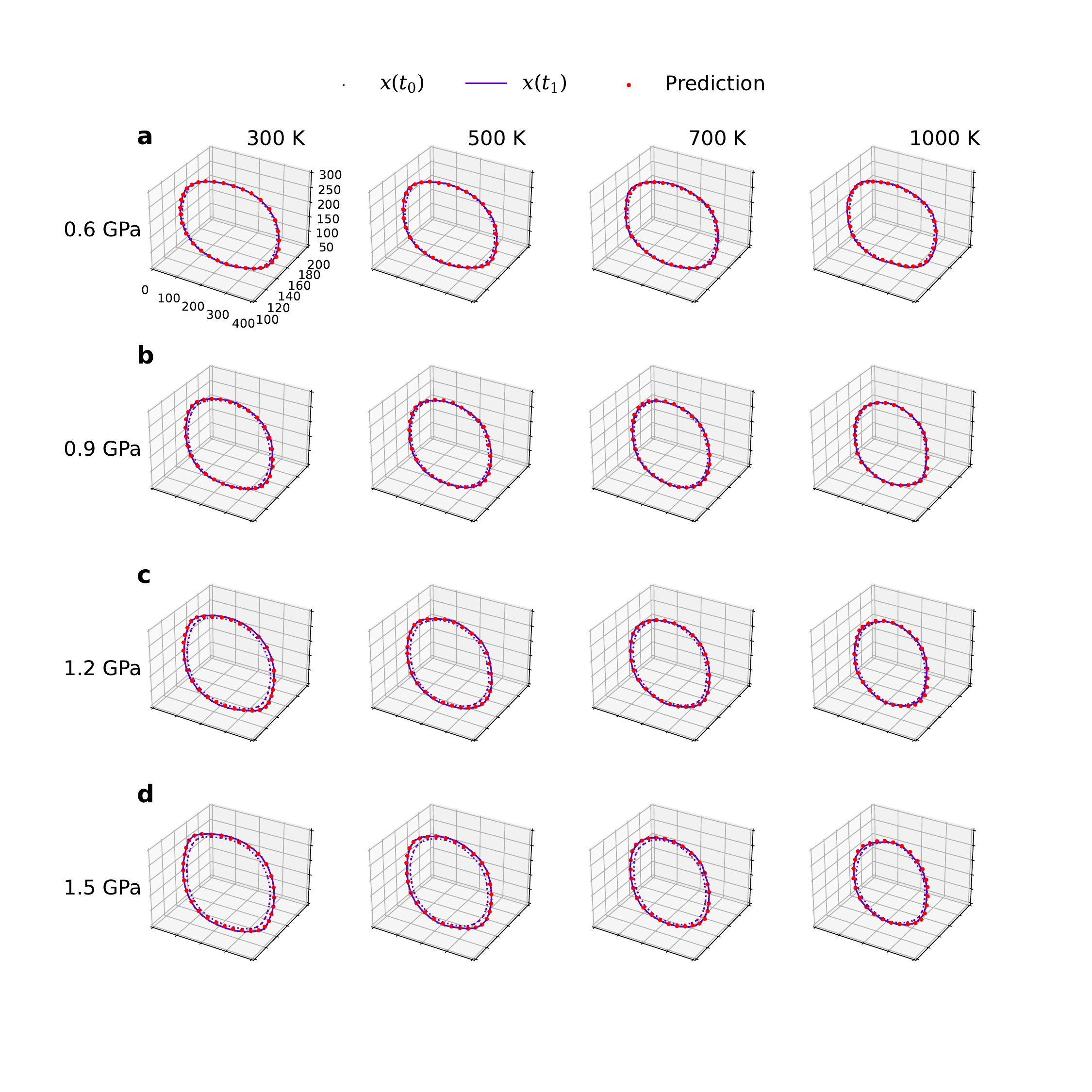}
    \caption{ Additional results for the predicted single-step evolution using PI-GNN mobility law trained on dislocation loop expansion data.}
    \label{fig:EvolutionLoopSup}
\end{figure}

\begin{figure}[htb]
    \centering\includegraphics[trim=0 240 0 240, width=0.9\textwidth]{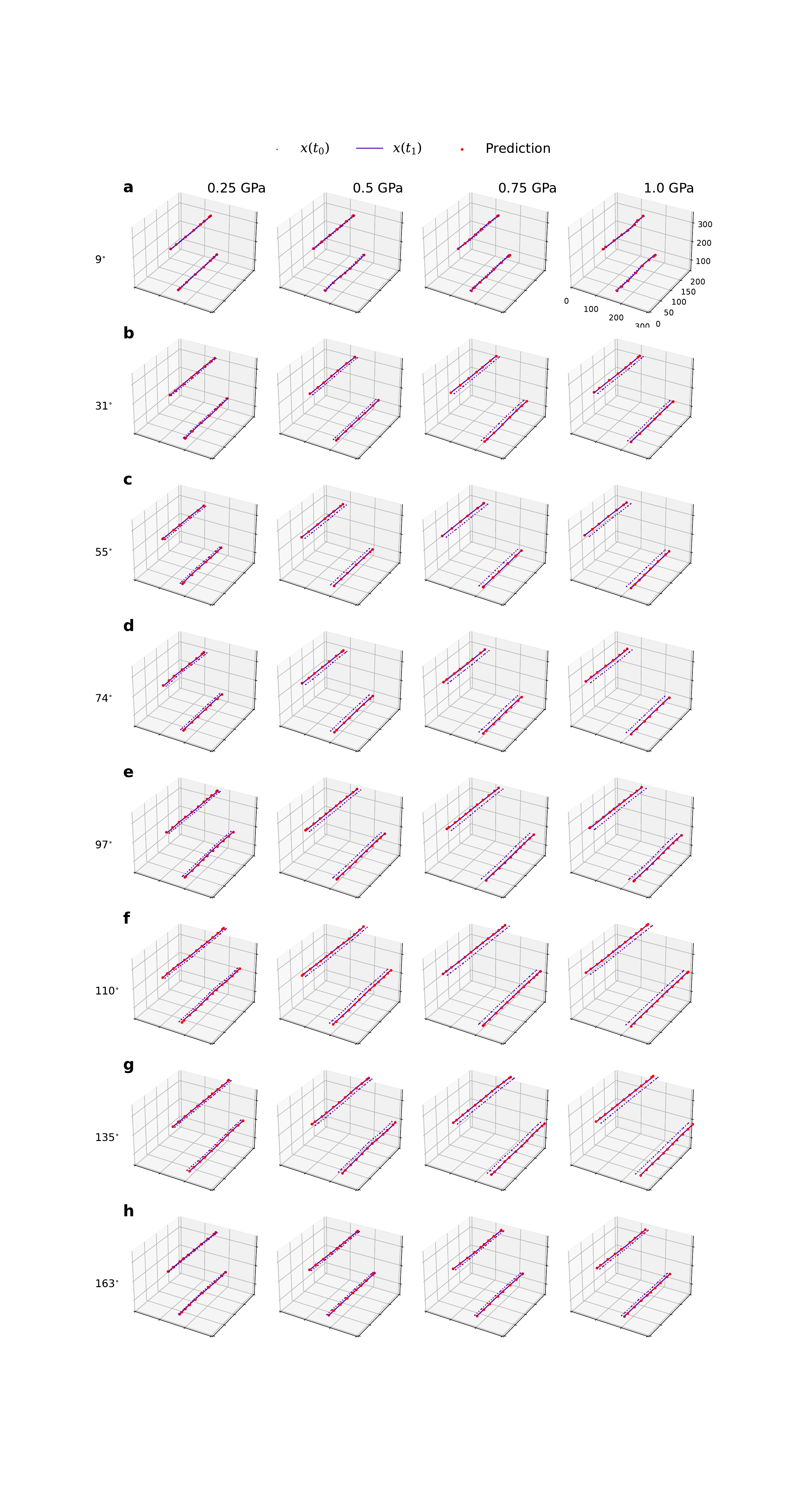}
    \caption{ Additional results for the predicted single-step evolution using PI-GNN mobility law trained on dislocation dipole data for different RSS and orientations.}
    \label{fig:EvolutionDipoleRSSSup}
\end{figure}

\begin{figure}[htb]
    \centering
    \includegraphics[trim=0 240 0 240, width=0.9\textwidth]{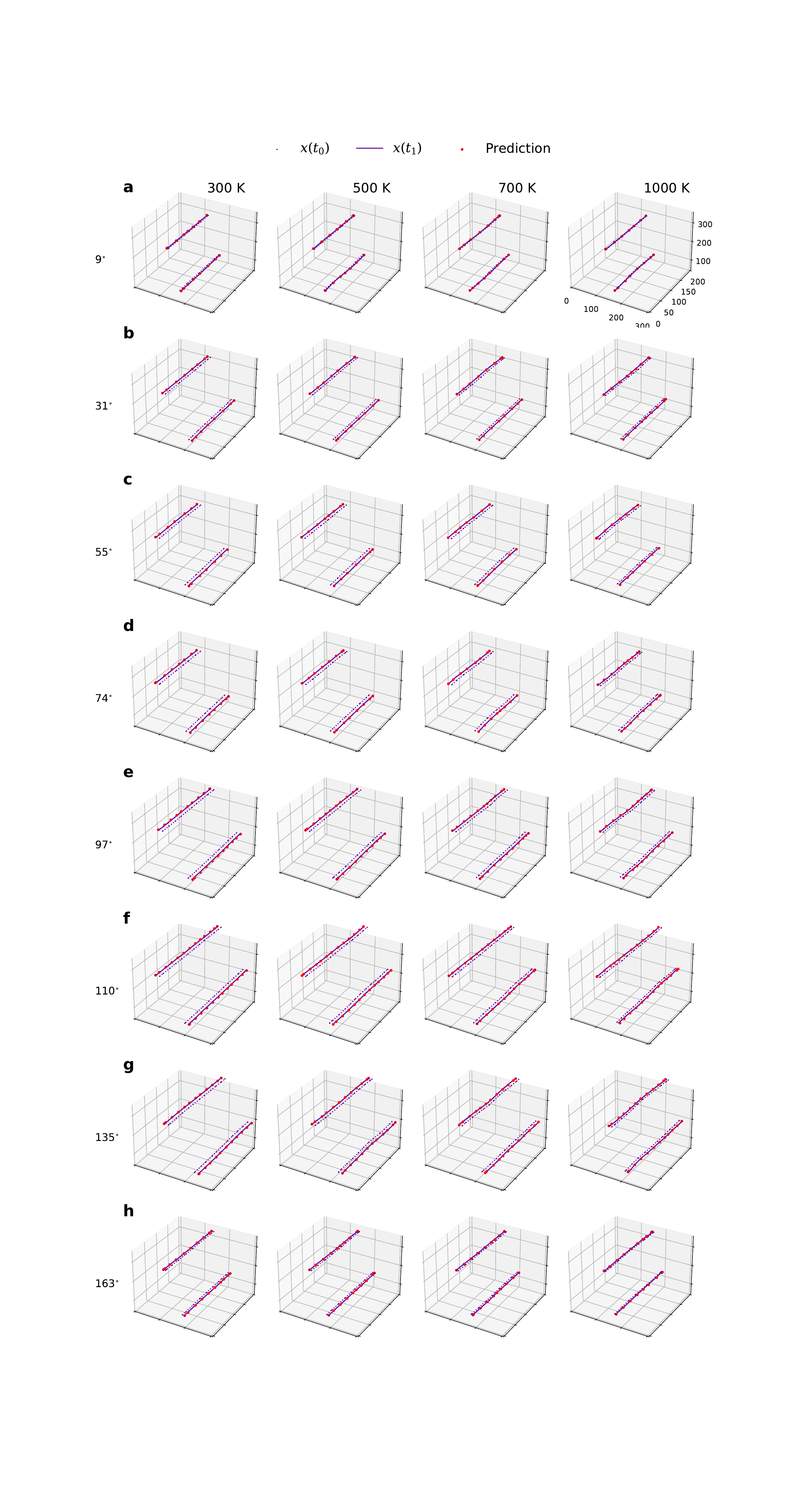}
    \caption{ Additional results for the predicted single-step evolution using PI-GNN mobility law trained on dislocation dipole data for different temperatures and orientations.}
    \label{fig:EvolutionDipoleTSup}
\end{figure}

In this section, we present additional results showing the prediction of dislocation evolution of the PI-GNN mobility law spanning a wider range of simulation parameters than those in the main text. The same training and prediction methods as in the main text are used. In Fig. \ref{fig:EvolutionLoopSup}, we present additional results for the single-step prediction of dislocation loops for various resolved shear stress (RSS) and temperatures. Over different configurations, we observe qualitatively good predictions by using the PI-GNN model. Similar additional results for dislocation dipoles are also shown in Figs. \ref{fig:EvolutionDipoleRSSSup} for different RSS levels and in Figs. \ref{fig:EvolutionDipoleTSup} for different temperatures.

\begin{figure}[htb]
    \centering
    \includegraphics[width=0.8\textwidth]{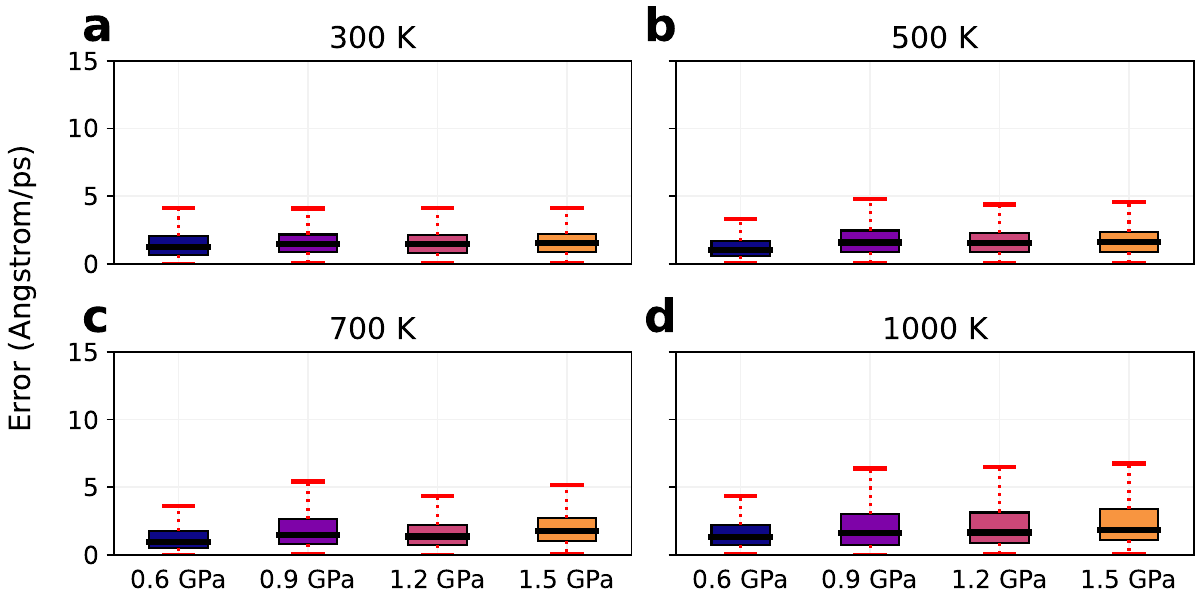}
    \caption{The normalized error of single-step prediction for dislocation loops under various RSS and temperatures.}
    \label{fig:l2EvolutionUnnormLoop}
\end{figure}

\begin{figure}[htb]
    \centering
    \includegraphics[width=0.8\textwidth]{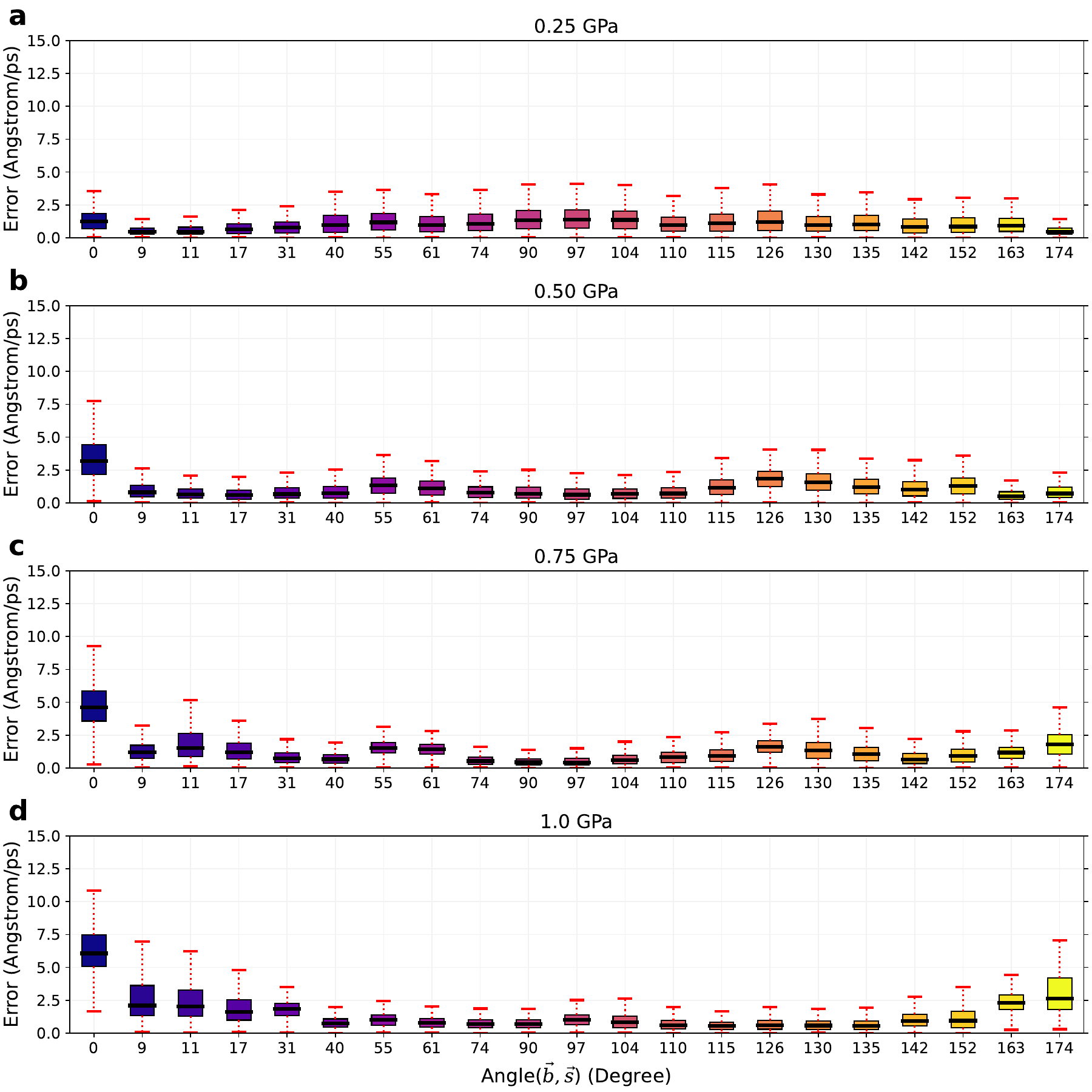}
    \caption{The absolute error of single-step prediction for dislocation dipoles under various resolved shear stress at 500K.}
    \label{fig:l2EvolutionUnnormDipoleRSS}
\end{figure}

\begin{figure}[htb]
    \centering
    \includegraphics[width=0.8\textwidth]{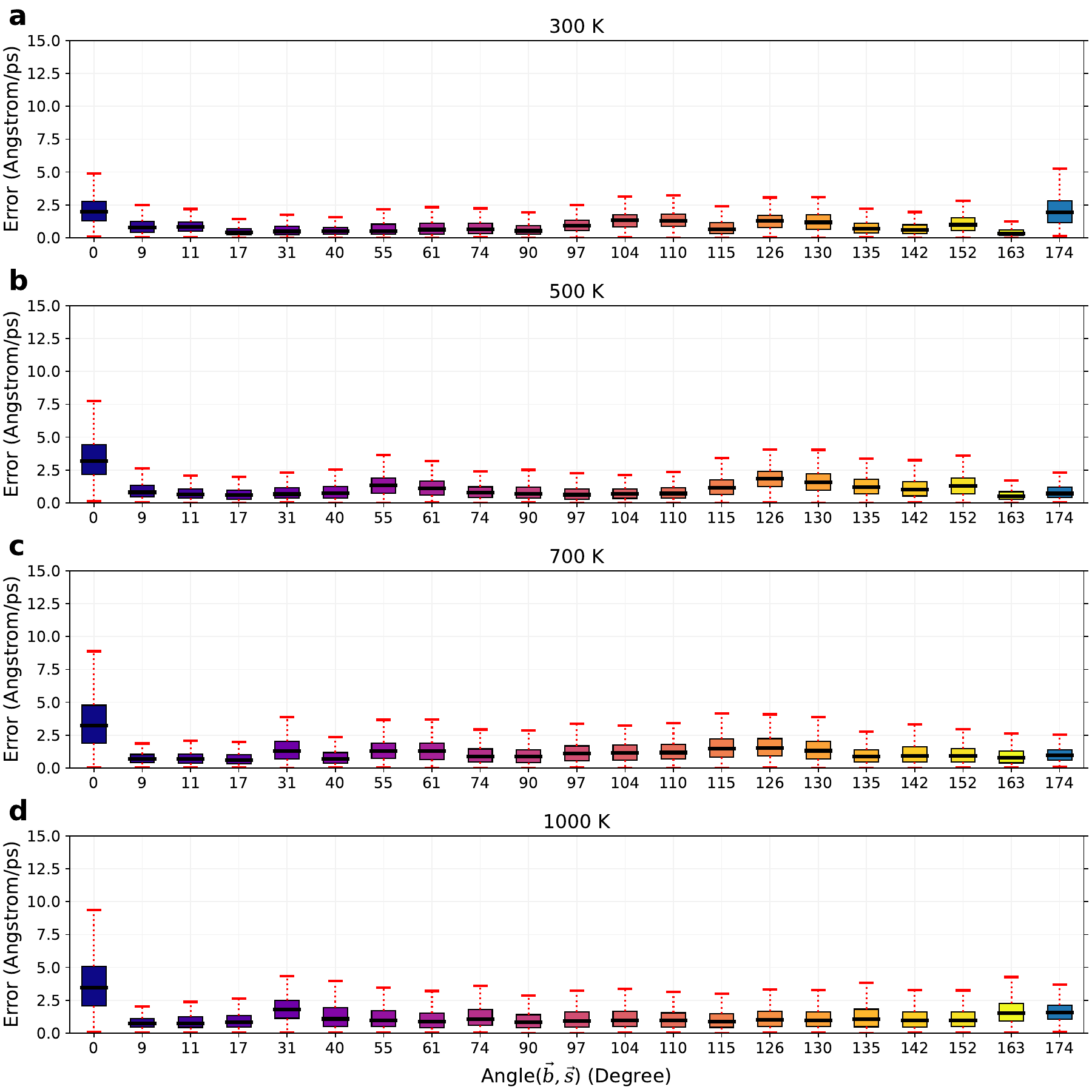}
    \caption{The absolute error of single-step prediction for dislocation dipoles under various temperatures and resolved shear stress of 0.5GPa.}
    \label{fig:l2EvolutionUnnormDipoleT}
\end{figure}

\begin{figure}[htb]
    \centering
    \includegraphics[width=0.8\textwidth]{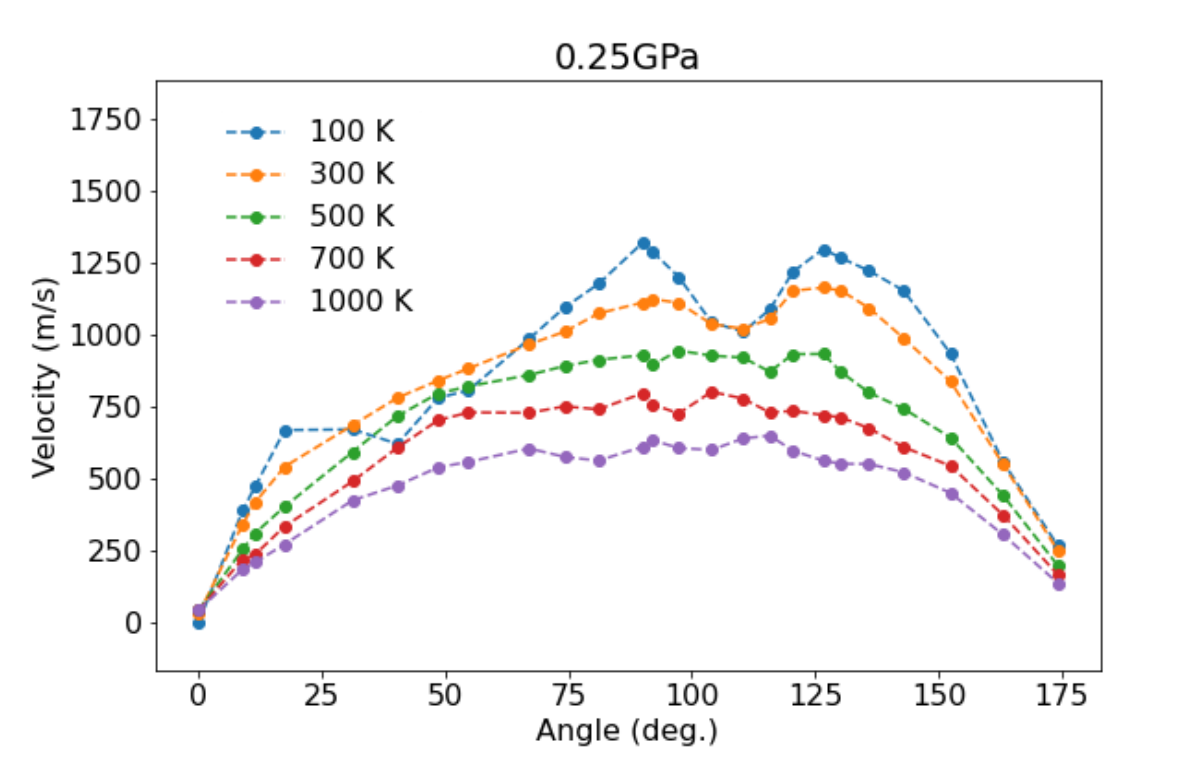}
    \caption{Orientation dependence of mobility for various temperatures, at a resolved shear stress of $0.25$ GPa, predicted from center of mass based trajectory analysis of dislocation lines from raw MD data. The cusp at $110^\circ$ is clearly visible for the lower temperatures.}
    \label{fig:rawMDcomMobility0.25GPa}
\end{figure}

\begin{figure}[htb]
    \centering
    \includegraphics[width=\textwidth]{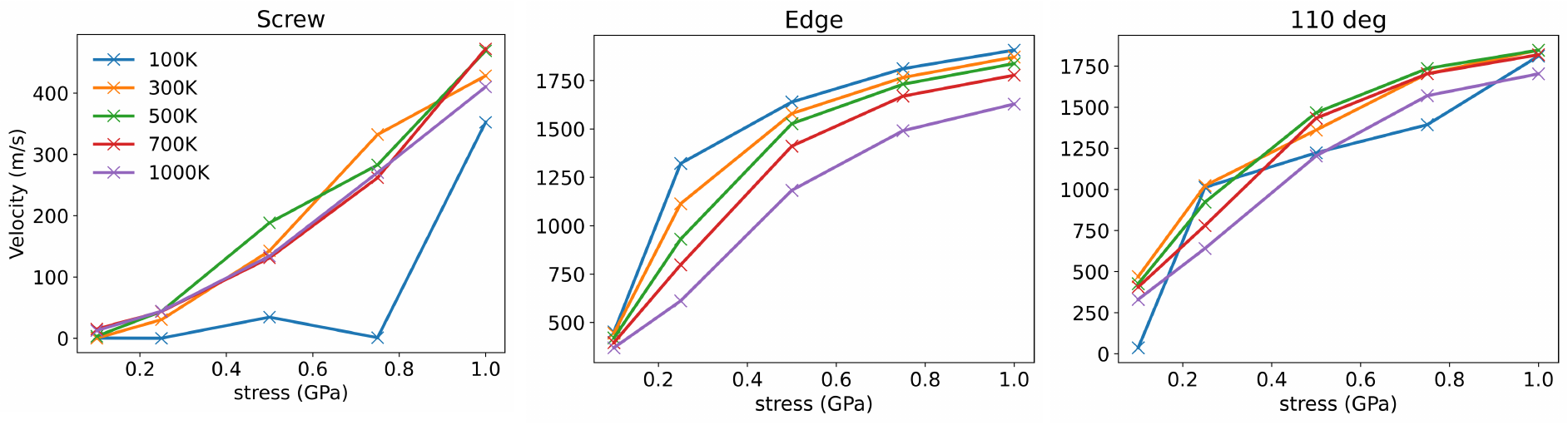}
    \caption{Velocity as a function of stress and temperature for Screw, Edge and M111 ($110^\circ$) characters of dislocation lines center of mass based trajectory analysis from raw MD data}
    \label{fig:Schew_EdgeM111Mobility}
\end{figure}

Quantitative analyses of the single-step prediction error are shown in Figs. \ref{fig:l2EvolutionUnnormLoop}, \ref{fig:l2EvolutionUnnormDipoleRSS}, \ref{fig:l2EvolutionUnnormDipoleT}. Note that the errors are in their original unit (Angstorm/ps) and not normalized. 
To validate our PI-GNN predictions, results from the traditional center of mass based mobility analysis \cite{gilbert2011stress,maresca2018screw,queyreau2011edge} of dislocation lines from raw MD data is presented in Fig. \ref{fig:rawMDcomMobility0.25GPa}. We compute the dislocation displacements from time-series of center of mass motion computed from DXA-predicted nodal positions belonging to the line configurations in snapshots with a frequency of $100$ fs. To ensure steady-state motion of dislocations, the first $5$ ps of the trajectory are discarded. Velocities are then computed using linear fits of displacement-time profiles. We plot velocities as function of orientation ($\vec{b}, \vec{s}$) and temperatures at RSS=$0.25$ GPa for mixed dipoles averaged over the dislocation pairs with positive and negative Burgers vectors in Fig. \ref{fig:rawMDcomMobility0.25GPa}. Velocities as a function of both stress and temperature is also shown for special line orientations e.g., screw, edge and the singular M111 ($110^\circ$) in Fig. \ref{fig:Schew_EdgeM111Mobility}. The ranges of these velocities follow similar trends as reported in previous analysis for edge \cite{queyreau2011edge} and screw \cite{gilbert2011stress} for the Mendelev \cite{mendelev2003development} potential. We note that the qualitative trend of orientation dependence revealed by the raw mobility data from MD is well captured by our PI-GNN model predictions as shown in the main text.

\subsection{Additional results on the multi-step prediction quality of the learned PI-GNN mobility law}

\begin{figure}[htb]
    \centering
    \includegraphics[width=0.9\textwidth]{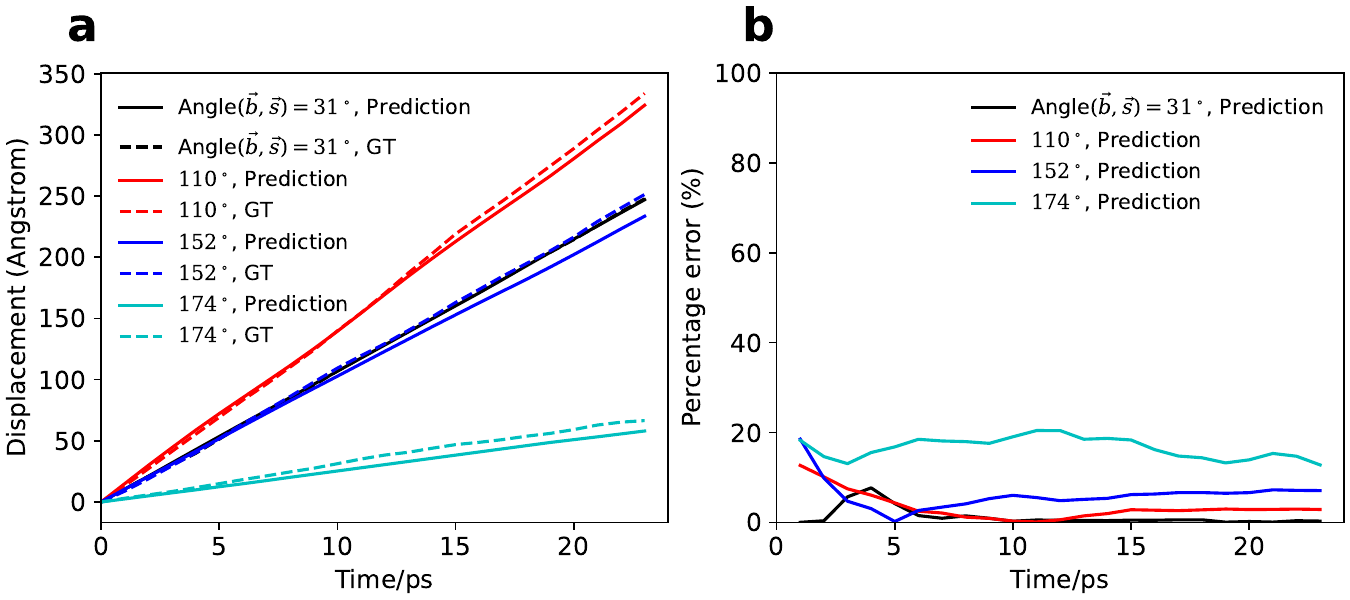}
    \caption{ Additional results for the predicted multi-step evolution using PI-GNN mobility law trained on dislocation dipole data for different orientations at 500K and 0.5GPa.}
    \label{fig:MultistepError}
\end{figure}

Here we provide additional results for multi-step predictions using the learned PI-GNN mobility law. The multi-step predictions are enabled by coupling the learned mobility law with the ModeLib framework for feature computation. In Fig. \ref{fig:MultistepError}, we demonstrate the accuracy and stability of the PI-GNN mobility law for the dislocation dipole configurations. We select three dislocation dipole cases with $\text{Angle}({\bf s}, {\bf b}) \in \{ 31^\circ, 110^\circ, 152^\circ, 174^\circ\}, $ at 500K and 0.5GPa. In Fig. \ref{fig:MultistepError} (a), we compare the displacement for the center of mass of the dislocation dipoles between the PI-GNN predictions and GT extracted from MD. We observe that the predictions remain stable for the learned PI-GNN model and agree qualitatively well with the GT. Towards the end of the prediction, the predictions slowly deviates from the GT due to the accumulation of errors, which is usually expected for reduced order models. In Fig. \ref{fig:MultistepError} (b), we demonstrate that the percentage errors of long-time predictions remain less than 20\%.

\clearpage
\newpage

\bibliographystyle{unsrt}
\bibliography{refs_arxiv}